\documentclass[aps,prd,preprintnumbers,twocolumn,groupedaddress,nofootinbib,showpacs]{revtex4-1}
\usepackage{graphicx}
\usepackage{amsmath}
\usepackage{amssymb}
\usepackage{mathtools}
\usepackage{subfigure}
\usepackage{float} 

\usepackage[final]{pdfpages}
     \DeclareMathOperator{\cm}{cm}

\newcommand{\pL}{\left(} \newcommand{\pR}{\right)} \newcommand{\bL}{\left[} \newcommand{\bR}{\right]}    
\newcommand{\beq}{\begin{equation}} \newcommand{\eeq}{\end{equation}}
\newcommand{\bea}{\begin{eqnarray}} \newcommand{\eea}{\end{eqnarray}}

\newcommand{\Eq}[1]{Eq.~(\ref{#1})}  
  
\newcommand{\Fig}[1]{Fig.~\ref{#1}}

\def\lsim{\mathrel{\raise.3ex\hbox{$<$\kern-.75em\lower1ex\hbox{$\sim$}}}}
\def\gsim{\mathrel{\raise.3ex\hbox{$>$\kern-.75em\lower1ex\hbox{$\sim$}}}}

\begin{document}

\title{Dissecting the Gamma-Ray Background in Search of Dark Matter }  
\author{Ilias Cholis$^{1}$}
\author{Dan Hooper$^{1,2}$}
\author{Samuel D.~McDermott$^{1,3}$}
\affiliation{$^1$Center for Particle Astrophysics, Fermi National Accelerator Laboratory, Batavia, IL 60510}
\affiliation{$^2$Department of Astronomy and Astrophysics, University of Chicago, Chicago, IL 60637}
\affiliation{$^3$Michigan Center for Theoretical Physics, University of Michigan, Ann Arbor, MI 48105}
\date{\today}

\begin{abstract}

Several classes of astrophysical sources contribute to the approximately isotropic gamma-ray background measured by the Fermi Gamma-Ray Space Telescope. In this paper, we use Fermi's catalog of gamma-ray sources (along with corresponding source catalogs at infrared and radio wavelengths) to build and constrain a model for the contributions to the extragalactic gamma-ray background from astrophysical sources, including radio galaxies, star-forming galaxies, and blazars. We then combine our model with Fermi's measurement of the gamma-ray background to derive constraints on the dark matter annihilation cross section, including contributions from both extragalactic and galactic halos and subhalos. The resulting constraints are competitive with the strongest current constraints from the Galactic Center and dwarf spheroidal galaxies. As Fermi continues to measure the gamma-ray emission from a greater number of astrophysical sources, it will become possible to more tightly constrain the astrophysical contributions to the extragalactic gamma-ray background. We project that with 10 years of data, Fermi's measurement of this background combined with the improved constraints on the astrophysical source contributions will yield a sensitivity to dark matter annihilations that exceeds the strongest current constraints by a factor of $\sim$5-10.

\end{abstract}

\pacs{95.35.+d, 07.85.-m, FERMILAB-PUB-13-546-A, MCTP-13-40}
\maketitle

\section{Introduction}

The diffuse and approximately isotropic gamma-ray background was first detected by the SAS-2 satellite~\cite{sas}, and later confirmed by EGRET~\cite{Sreekumar:1997un} and the Fermi Gamma-Ray Space Telescope~\cite{Abdo:2010nz}. This emission has long been speculated to be the product of a large number of unresolved sources, such as active galactic nuclei~\cite{Stecker:1993ni,Padovani, Salamon:1994ku,Stecker:1996ma,Mukherjee:1999it,Narumoto:2006qg,Giommi:2005bp,Dermer:2006pd,Pavlidou:2007dv,Inoue:2008pk} or star-forming galaxies~\cite{Pavlidou:2002va,Thompson:2006qd,Fields:2010bw,Makiya:2010zt}. It was also suggested that a portion of this background could be the result of annihilating dark matter particles~\cite{Stecker:1978du,Gao:1991rz,Ullio:2002pj}. 

With the wealth of new information brought forth by Fermi, a much more concrete and detailed picture for the origin of the extragalactic gamma-ray background (EGB) has emerged.\footnote{Although we will use the phrase ``extragalactic gamma-ray background'' throughout this paper to describe the approximately isotropic emission that is observed, we do not intend to imply that no galactic sources could contribute to this flux. If distributed sufficiently isotropically across the sky, a population of faint galactic sources would be difficult to separate from the extragalactic background. Despite the recent detection of small scale anisotropies~\cite{SiegalGaskins:2010nh}, this background is also  sometimes referred to as the isotropic gamma-ray background.} In particular, the large catalog of blazars observed by Fermi~\cite{Ackermann2011} has been used to construct detailed luminosity functions and redshift distributions for the populations of flat-spectrum radio quasars (FSRQs) and BL Lac objects. This information, as well as the degree of small-scale anisotropy observed by Fermi~\cite{SiegalGaskins:2010nh}, supports the conclusion that unresolved blazars contribute only $\sim$20\% of the EGB~\cite{Cuoco:2012yf,Harding:2012gk,Collaboration:2010gqa,Ajello:2011zi}. Fermi's detection of gamma-ray emission from both star-forming galaxies~\cite{Ackermann:2012vca} and radio galaxies~\cite{Inoue:2011bm}, combined with the observed correlations between emission at gamma-ray and infrared and radio wavelengths, has revealed that these source classes each contribute significantly to the EGB. Taken together, the emission from unresolved blazars, star-forming galaxies, and radio galaxies is likely to make up the majority of the observed EGB, and could plausibly constitute the entirety of this background (see, for example, the  combinations presented in Refs.~\cite{Siegal-Gaskins:2013tga} and~\cite{Cavadini:2011ig}, or the discussion in Ref.~\cite{Stecker:2010di}). Given the uncertainties in the characteristics of these source populations, however, there remains room for not-insignificant contributions from other sources, such as merging galaxy clusters~\cite{Keshet:2002sw,Gabici:2002fg,Gabici:2003kr}, cascades generated in the propagation of ultra-high energy cosmic rays~\cite{Ahlers:2011sd,Gelmini:2011kg}, or annihilating or decaying dark matter.

\begin{figure*}
\mbox{\includegraphics[width=0.49\textwidth,clip]{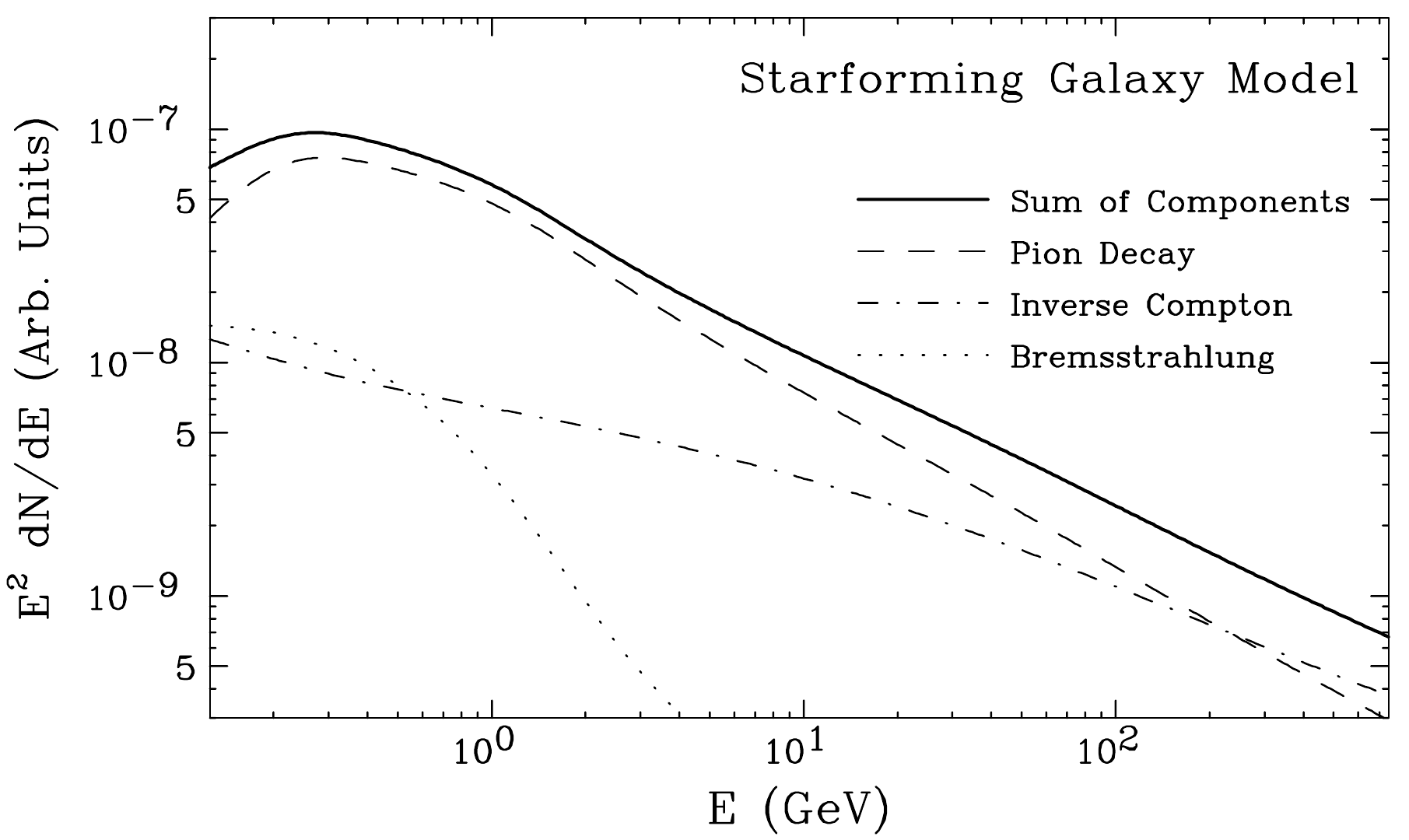}} 
\mbox{\includegraphics[width=0.49\textwidth,clip]{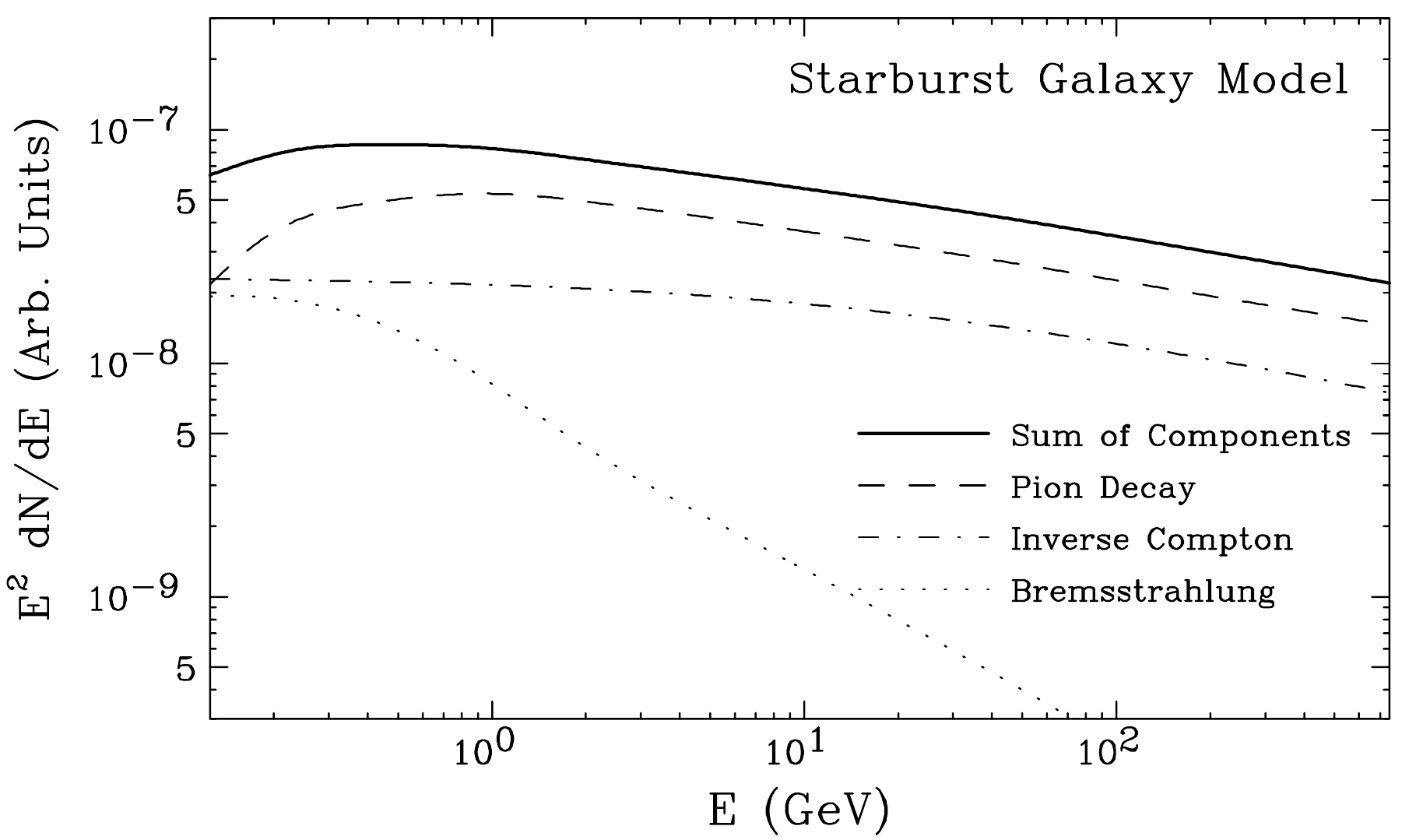}} 
\caption{The models used in our analysis to describe the spectral shape of the gamma-ray emission from Milky Way-like star-forming galaxies (left) and much higher luminosity starburst galaxies (right), neglecting attenuation from the cosmic infrared background.  See text for details.}
\label{SF1}
\end{figure*}

In this paper, we construct an empirically based model for the contributions to the EGB from star-forming galaxies, radio galaxies, FSRQs, and BL Lac objects, and we compare this model to the observed spectrum of the EGB. We then make use of this model to derive upper limits on the contribution from dark matter, and on the corresponding annihilation cross sections.  We find that the resulting dark matter constraints are competitive with those derived from observations of the Galactic Center~\cite{Hooper:2012sr} and dwarf spheroidal galaxies~\cite{Ackermann:2013yva,Abramowski:2010aa}. Furthermore, as Fermi continues to detect and characterize the gamma-ray emission from an ever larger number of sources, it will become increasingly possible to tightly constrain the various astrophysical contributions to the EGB. We project that with 10 years of data, Fermi's measurement of the EGB, combined with the expected constraints on the astrophysical source contributions, will yield a sensitivity to dark matter annihilations that exceeds current constraints by a factor of $\sim$5-10. Such a result could plausibly represent the strongest constraint on the dark matter annihilation cross section by the end of the Fermi mission.

The remainder of this paper is structured as follows. In Sec.~\ref{astro}, we discuss contributions to the EGB from a variety of astrophysical sources, including star-forming galaxies, radio galaxies, blazars, cascades induced by ultra-high energy cosmic rays, and millisecond pulsars. We describe and constrain a model for this astrophysical emission, and find that the combination of these sources could account for the entirety of the EGB, although with significant statistical and systematic uncertainties. In Sec.~\ref{dm}, we calculate the contribution to the EGB from dark matter annihilations, including extragalactic halos and subhalos, and the halo and subhalos of the Milky Way. In Sec.~\ref{constraints}, we use these results to derive constraints on the dark matter annihilation cross section. In Sec.~\ref{projections}, we make projections for Fermi's future sensitivity to annihilating dark matter. Finally, in Sec.~\ref{conclusions}, we summarize our results and conclusions.

\section{Astrophysical Contributions to the Diffuse Gamma-Ray Background}
\label{astro}

In this section, we discuss several astrophysical contributions to the EGB, and constrain their spectral shapes and normalizations. Taken together, we find that the combination of emission from star-forming galaxies, radio galaxies, FSRQs and BL Lac objects likely makes up the majority of the EGB, although with significant uncertainties.

\subsection{Star-Forming Galaxies}
\label{sfsec}

Although few galaxies (excluding those with active nuclei) are bright enough to be detected by Fermi as individual sources, they are very numerous and may collectively contribute significantly to the EGB~\cite{Thompson:2006qd,Fields:2010bw,Makiya:2010zt,Stecker:2010di}.  Galaxies produce and contain cosmic rays, which generate gamma-rays through pion decay, inverse Compton, and bremsstrahlung processes. The intensity and spectrum of this emission is expected to depend on the star formation history of the galaxy in question. To date, Fermi has reported the detection of only nine individual galaxies, four of which reside within the Local Group (the SMC, LMC, M31, and Milky Way) and five of which are more distant (NGC 253, M82, NGC 4945, NGC 1068 and Circinus)~\cite{Ackermann:2012vca,Abdo:2009aa,Lacki:2010vs,Hayashida:2013wha}. Additionally, M82 and NGC 253 have been observed at very high-energies by ground-based gamma-ray telescopes~\cite{veritasSF,hessSF}. Taking this information alone, it would be very difficult to produce a reliable model for the luminosity and redshift distribution of such sources. Fortunately, many more galaxies have been detected at infrared wavelengths~\cite{IRlumfunc}, and the gamma-ray luminosities of the galaxies detected by Fermi have been shown to be highly correlated with the corresponding radio and infrared emission. In particular, Ref.~\cite{Ackermann:2012vca} reports the following relationship between the emission in the 0.1-100 GeV and 8-1000 $\mu$m bands:
\begin{equation}
\log\bigg(\frac{L_{0.1-100 \, {\rm GeV}}}{{\rm erg/s}}\bigg) = \alpha \log\bigg(\frac{L_{8-1000\,\mu{\rm m}}}{10^{10} \, L_{\odot}}\bigg) + \beta,
\end{equation}
where $\alpha=1.17 \pm 0.07$ and $\beta=39.28 \pm 0.08$. Combining this observed correlation with the observed infrared luminosity function and redshift distribution of galaxies~\cite{IRlumfunc}, it is possible to derive the gamma-ray luminosity function for this source population~\cite{Ackermann:2012vca}. 

To describe the spectral shape from this source population, we build a physical model for the gamma-ray spectra from star-forming and starburst galaxies, constrained to match the observed emission from such objects. In the left and right frames of Fig.~\ref{SF1}, we plot the gamma-ray spectrum from a Milky Way-like star-forming galaxy and a high luminosity starburst galaxy, respectively. In the starburst case, we select the pion, inverse Compton, and bremsstrahlung components to match the overall spectral index (above 1 GeV) of 2.2, as observed from individual starburst galaxies by Fermi~\cite{Ackermann:2012vca}. In the Milky Way-like case, we normalized the various components (relatively) according to the model described in Ref.~\cite{Strong:2010pr}. In calculating the contribution to the diffuse gamma-ray background, we describe the spectral shape from the combination of all star-forming galaxies (including starburst galaxies) as a weighted sum which is a function of a single parameter, $f$: 
\begin{equation}
\frac{dN_{\gamma}}{dE_{\gamma}} = f  \, \frac{dN_{\gamma}}{dE_{\gamma}}\bigg|_{\rm star-forming} + (1-f) \, \frac{dN_{\gamma}}{dE_{\gamma}}\bigg|_{\rm starburst}. 
\end{equation}

\begin{figure}
\mbox{\includegraphics[width=0.49\textwidth,clip]{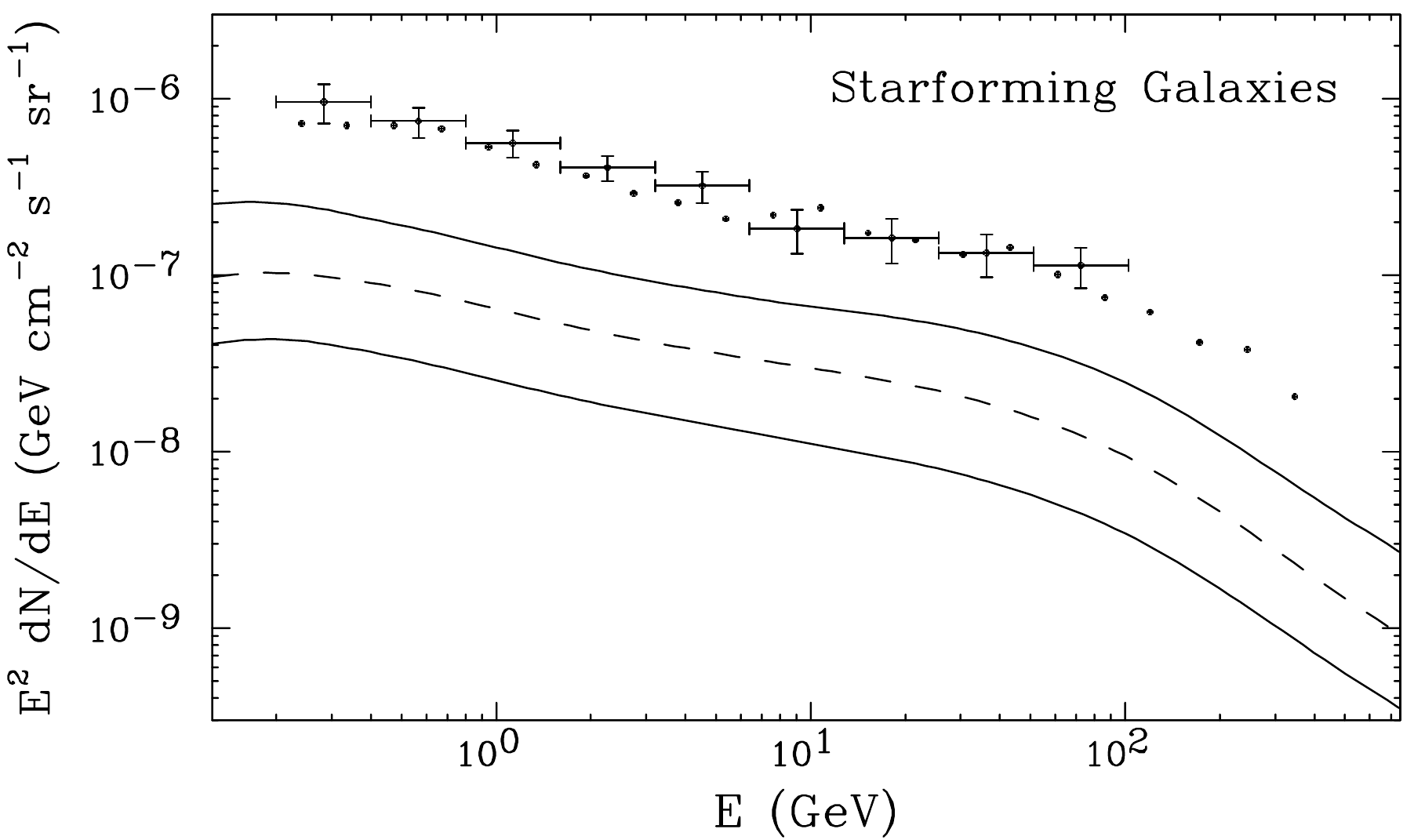}} 
\caption{The estimated contribution to the EGB from star-forming galaxies (including starburst galaxies). The dashed curve represents the estimate derived using the central parameter values, while the solid lines are the 1$\sigma$ uncertainties around that prediction. The error bars denote the spectrum of the EGB as measured by Fermi~\cite{Abdo:2010nz}, while the points without error bars are the central values of the Fermi's preliminary EGB analysis, currently in preparation and shown only for comparison~\cite{preliminaryfermi}. See text for details.}
\label{SF}
\end{figure}

In Fig.~\ref{SF}, we show our estimate for the contribution to the EGB from star-forming galaxies. The central (dashed) curve corresponds to the result found for $\alpha=1.17$, $\beta=39.28$, $f=0.5$, and the central value of the normalization of the infrared luminosity function~\cite{IRlumfunc}. To calculate the uncertainty for this contribution (solid, representing variations at the 1$\sigma$ level), we propagate the following uncertainties in these parameters: $\alpha=1.17 \pm 0.07$, $\beta=39.28 \pm 0.08$, $f=0.5 \times 10^{\pm0.20}$ (constrained such that $0 < f < 1$), and an overall uncertainty of $\pm$30\% in the normalization of the infrared luminosity function. Taken together, we find that while star-forming galaxies are likely to produce only $\sim$10-15\% of the extragalactic diffuse gamma-ray background, the related uncertainties are large, allowing for the possibility that their contribution could be more significant. Throughout this study, we adopt standard cosmological parameters ($\Omega_{\Lambda}=0.6817$, $\Omega_M=0.3183$~\cite{Ade:2013zuv}) and account for gamma-ray attenuation via pair-production with the cosmic infrared background ($\gamma + \gamma_{\rm IR} \rightarrow e^+ e^-$) using Ref.~\cite{Gilmore:2011ks}'s ``fiducial model'' for the optical depth, $\tau(E_{\gamma}, z)$.\footnote{The ``fiducial model'' of Ref.~\cite{Gilmore:2011ks} accounts for the evolution of the absorption efficiency of dust with redshift.  If we had instead adopted their ``fixed model'' for the optical depth, this would have impact our limits by a factor of $\sim$2 for dark matter masses greater than a few TeV, and insignificantly for masses below $\sim$500 GeV.}

\subsection{Radio Galaxies}
\label{rgsec}

\begin{figure}
\mbox{\includegraphics[width=0.49\textwidth,clip]{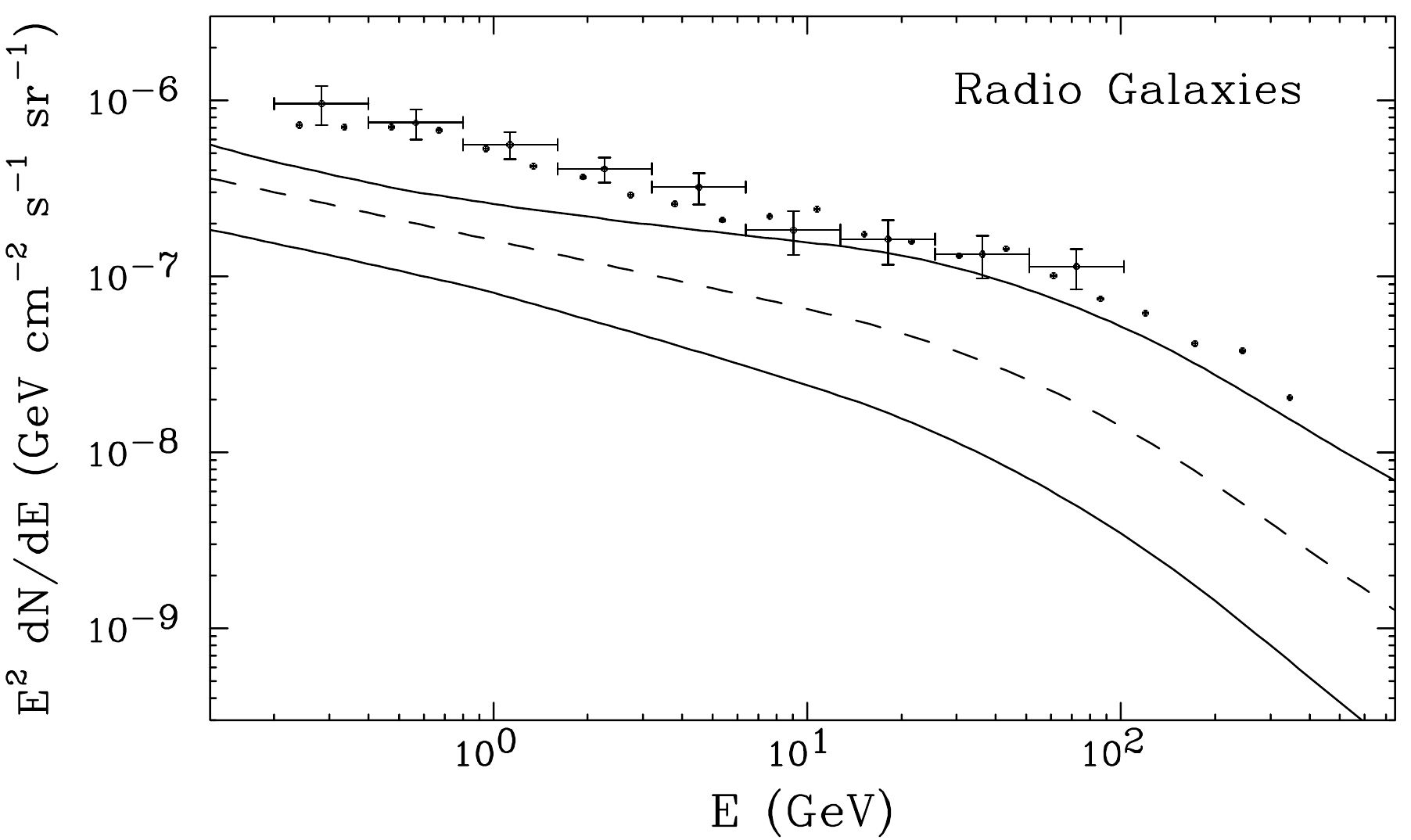}} 
\caption{The estimated contribution to the EGB from radio galaxies (including both FRI and FRII galaxies). The dashed curve represents the estimate derived using the central parameter values, while the solid lines are the 1$\sigma$ uncertainties around that prediction. Error bars and points are as in Fig.~\ref{SF}. See text for details.}
\label{RG}
\end{figure}

Radio galaxies are active galactic nuclei with relativistic jets that are not aligned with our line-of-sight. Within this context, Fanaroff-Riley (FR) type I and II radio galaxies are misaligned BL Lacs and FSRQs, respectively \cite{Urry:1995mg}. Although radio galaxies are much fainter than blazars, they are also much more numerous. As a result, gamma-ray emission from unresolved radio galaxies is expected to contribute significantly to the EGB~\cite{Inoue:2011bm,Stecker:2010di,DiMauro:2013xta}.

As with star-forming galaxies, only a small number (eleven at present) of radio galaxies have been detected at GeV energies~\cite{Fermi11RG}. But also like star-forming galaxies, a strong correlation has been observed between the GeV emission of radio galaxies and the emission produced at other wavelengths. In particular, the gamma-ray (0.1-10 GeV) and radio (5 GHz) emission from both FRI and FRII radio galaxies exhibit the following correlation~\cite{Inoue:2011bm}:
\begin{equation}
\log\bigg(\frac{L_{0.1-10 \, {\rm GeV}}}{{\rm erg/s}}\bigg) = A \log\bigg(\frac{L_{5\,{\rm GHz}}}{{\rm erg/s}}\bigg) + B,
\end{equation}
where $A=1.16 \pm 0.02$ and $B=-3.90\pm 0.61$. We combine this observed correlation with the luminosity function and redshift distribution of radio galaxies as reported by Willott {\it et al.}~\cite{Willott:2000dh} (which includes both FRI and FRII type galaxies) to generate a model for the resulting gamma-ray emission. 

For the spectral shape of the gamma-ray emission from radio galaxies, we adopt a power-law with an index which we allow to vary from source-to-source around an average value. Using the ten spectral indices reported in Ref.~\cite{Inoue:2011bm}, we find that a good fit is found for an average spectral index of $\Gamma=2.39 \pm 0.15$, with a source-to-source variation of $\sigma \approx 0.2$. In addition to the uncertainties on the spectral index and the radio-GeV correlation parameters, we include a 14\% uncertainty in the overall normalization (corresponding to $\kappa=0.081 \pm 0.011$ in Ref.~\cite{Inoue:2011bm}). In Fig.~\ref{RG}, we show our estimate for the contribution of radio galaxies to the EGB, including the result derived using our central parameter values and the surrounding $1\sigma$ uncertainty band.

\subsection{Blazars}
\label{blazarsec}

Blazars are by far the most numerous class of resolved extragalactic gamma-ray sources, and were long considered to be a leading candidate to generate the majority of the EGB. As the number of detected sources increased, however, it became apparent that unresolved blazars are unlikely to dominate this background. Taken together with the observed degree of anisotropy in the diffuse gamma-ray background at high-latitudes~\cite{SiegalGaskins:2010nh}, blazars appear likely to account for only approximately 20\% of the EGB~\cite{Cuoco:2012yf,Harding:2012gk,Collaboration:2010gqa,Ajello:2011zi}.

\begin{figure}
\mbox{\includegraphics[width=0.49\textwidth,clip]{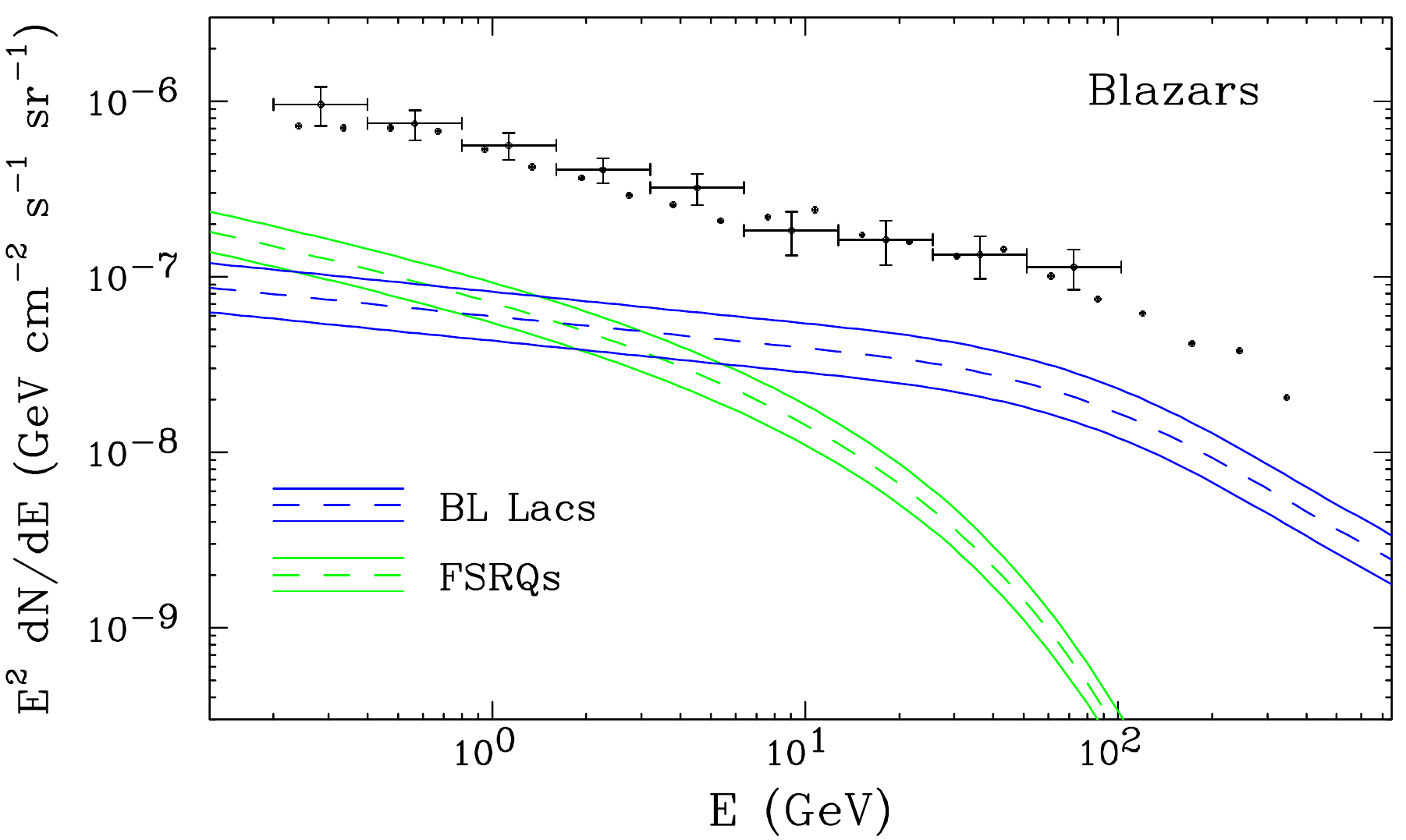}} 
\caption{The estimated contribution to the EGB from blazars (BL Lacs and FSRQs). Once again, the dashed curve represents the estimate derived using the central parameter values, while the solid lines are the 1$\sigma$ uncertainties around that prediction. Error bars and points are as in Fig.~\ref{SF}. See text for details.}
\label{blazars}
\end{figure}

To estimate the contribution from blazars to the EGB, we consider BL Lac objects and FSRQs independently. For each of these source classes, Fermi has resolved a large number of individual objects, making it possible to construct fairly reliable distributions of these sources in luminosity and redshift, without relying on correlations with emission at other wavelengths. Our method to estimate these contributions follows closely the works of Refs.~\cite{Ajello:2013lka,Collaboration:2010gqa,Ajello:2011zi}, and we do not repeat the details here. In Fig.~\ref{blazars}, we show the resulting contributions of BL Lacs and FSRQs to the EGB.

\begin{figure}
\mbox{\includegraphics[width=0.49\textwidth,clip]{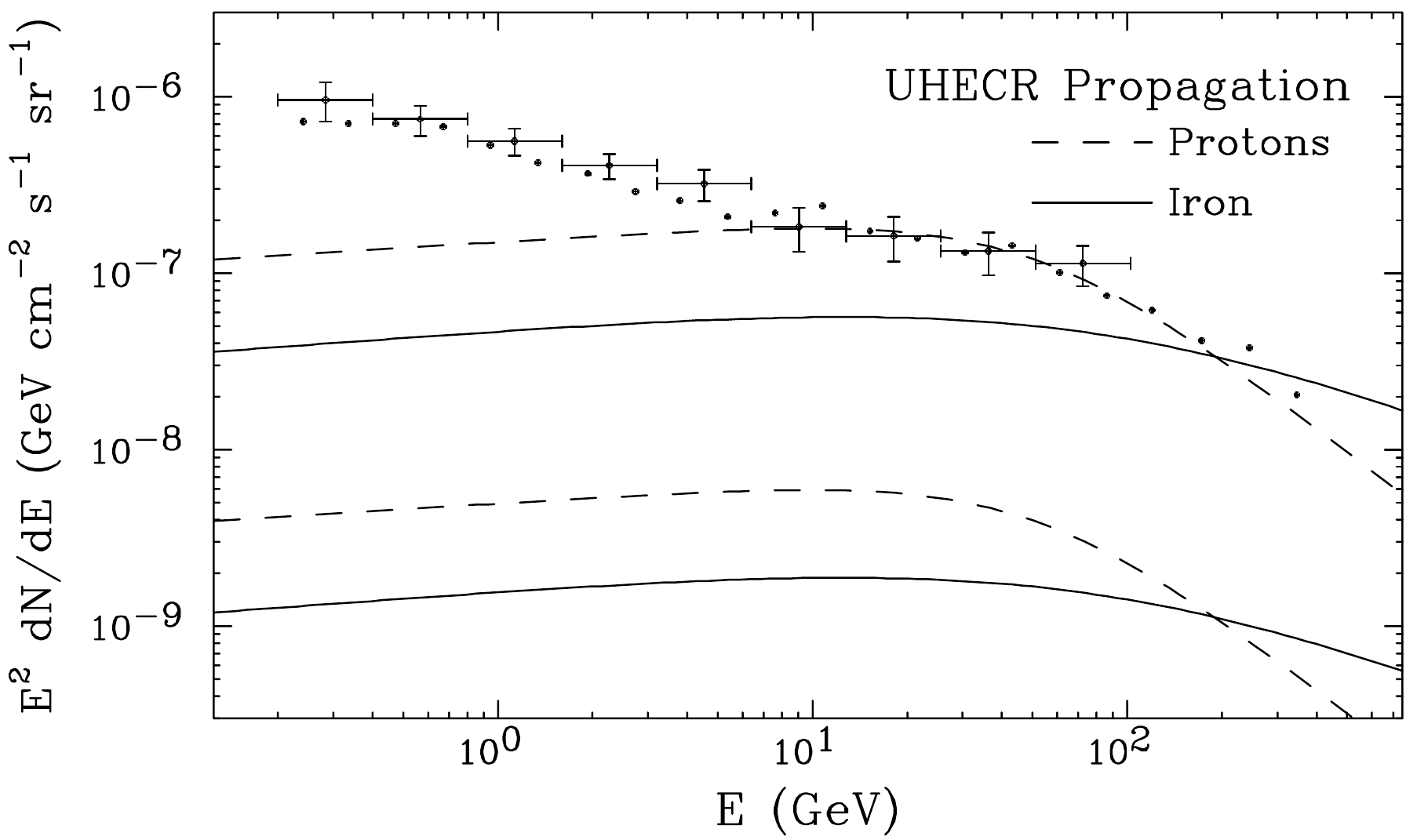}} 
\caption{The estimated contribution to the EGB from the propagation of ultra-high energy cosmic rays. The upper and lower sets of curves correspond to models with very strong source evolution and no source evolution, respectively~\cite{Ahlers:2011sd}. Error bars and points are as in Fig.~\ref{SF}. See text for details.}  
\label{uhecr}
\end{figure}

\subsection{Ultra-High Energy Cosmic Ray Propagation}

Ultra-high energy protons and nuclei scatter with the cosmic microwave and infrared backgrounds, leading to their attenuation and to the corresponding spectral feature known as the GZK cutoff~\cite{Greisen:1966jv,Zatsepin:1966jv}. Such interactions also initiate electromagnetic cascades.  The energetic photons and electrons associated with such cascades undergo a rapid sequence of pair production and inverse Compton scattering events, evolving rapidly downward in energy. The resulting spectrum of diffuse gamma-rays peaks at energies of $\sim$$10-100$ GeV, representing the approximate energy below which the universe is transparent to gamma-rays. 

The spectrum of gamma-rays resulting from ultra-high energy cosmic ray (UHECR) propagation depends on a number of relatively unconstrained factors, including the redshift distribution of sources, the chemical composition of the UHECRs, the extragalactic magnetic field distribution, and the energy density of the cosmic radio background. As a result, very large uncertainties are associated with the overall flux of gamma-rays produced by such particles. The spectral shape of this contribution, in contrast, is less sensitive to these unknown factors. In Fig.~\ref{uhecr} we show the contribution from UHECR propagation to the EGB for a few representative cases, as originally presented in Ref.~\cite{Ahlers:2011sd} (see also, Ref.~\cite{Gelmini:2011kg}). For each of the four curves shown, the injected cosmic ray spectrum is taken to consist purely of protons or iron nuclei, with a spectral index of 2.3, and with an exponential cutoff above $Z \times 10^{20.5}$ eV (where $Z=1$ for protons and 26 iron nuclei). The upper two curves assume a very strong source evolution, $n(z) = n_0 \, (1+z)^5$, while the lower two curves adopt an unchanging source distribution with redshift, $n(z)=n_0$. These cases shown are rather extreme, and the true contribution from UHECR propagation is likely to fall somewhere within this range.

\subsection{Millisecond Pulsars}

Pulsars are rapidly spinning neutron stars which steadily convert their rotational kinetic energy into radiation, including potentially observable emission at radio and gamma-ray wavelengths. Due to their long lifetimes and expected spatial distribution, unresolved millisecond pulsars (MSPs), also known as recycled pulsars, have been considered as potential contributors to the high-latitude diffuse gamma-ray background~\cite{loeb} (see also Ref.~\cite{averagepulsar}). 

The Fermi Collaboration has detected gamma-ray emission from a total of 125 sources identified as pulsars, 47 of which have millisecond-scale periods~\cite{publiclist}. Following Ref.~\cite{msp}, we build a spatial distribution and luminosity function model for galactic millisecond pulsars, constrained to account for the MSPs observed by Fermi without exceeding the total number of observed MSPs and currently unidentified gamma-ray sources. We also further constrain the spatial distribution to accommodate the distribution of such sources observed at radio frequencies~\cite{atnf}. Taken together, we find that MSPs are expected to account for only approximately 0.1\% to 0.3\% of the diffuse gamma-ray background above 1 GeV. This estimate is also compatible with constraints from Fermi's anisotropy measurement~\cite{SiegalGaskins:2010mp}. For details of the model used and its fit the observed MSP distribution, we direct the reader to Ref.~\cite{msp}. For the spectral shape of the gamma-ray emission from unresolved MSPs, we adopt $dN_{\gamma}/dE_{\gamma} \propto E_{\gamma}^{-1.46} \exp(-E_{\gamma}/3.3 \, {\rm GeV})$, which provides a good fit to the spectra observed from individual MSPs~\cite{msp}.

\begin{figure}
\mbox{\includegraphics[width=0.49\textwidth,clip]{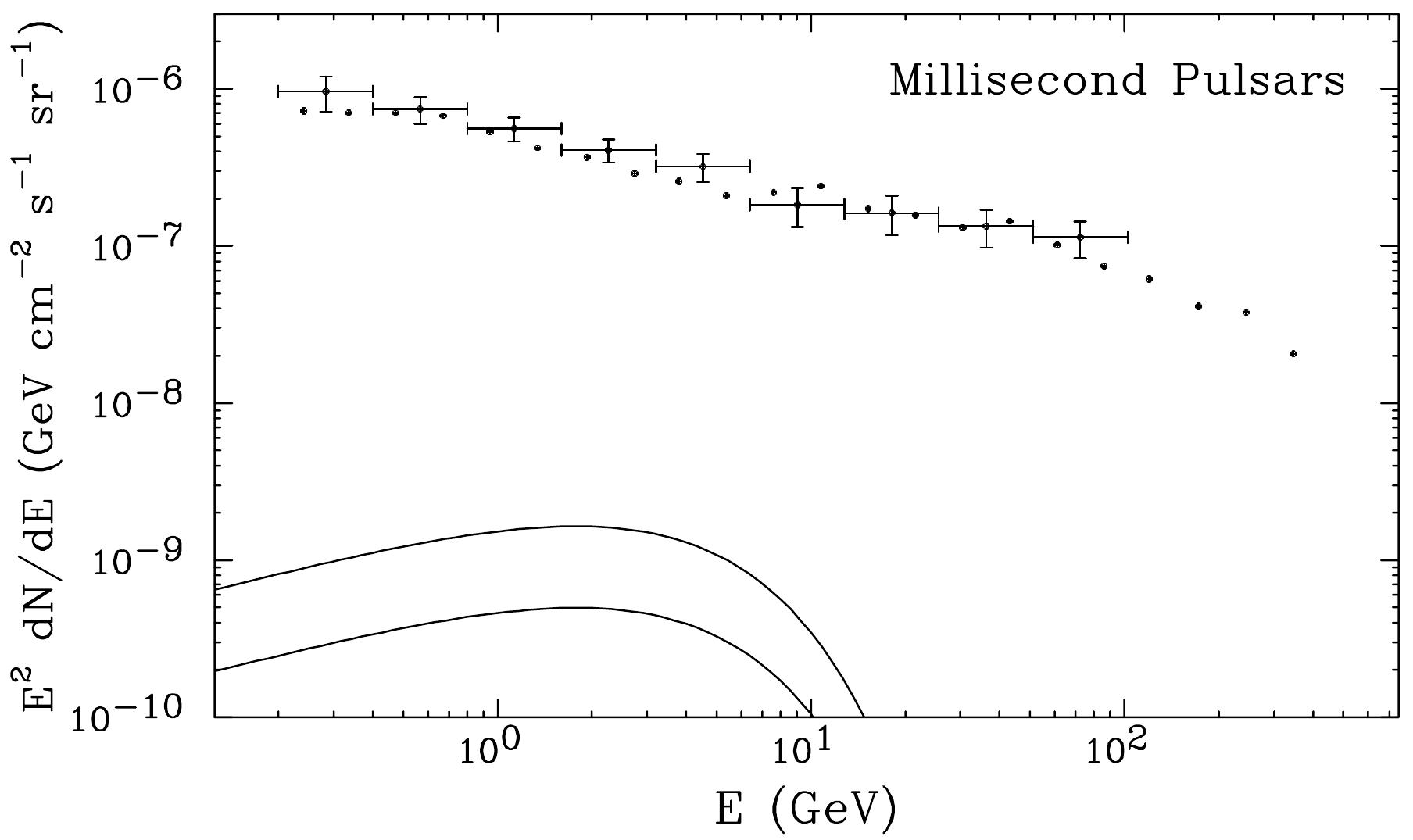}} 
\caption{The estimated contribution to the high-latitude, diffuse gamma-ray background from millisecond pulsars.  See text for details.}
\label{mspfig}
\end{figure}

In Fig.~\ref{mspfig}, we show our estimate for the contribution from millisecond pulsars to the diffuse gamma-ray background (integrated above $|b| >30^{\circ}$). The contribution has a negligible impact on our fits and limits, and thus we do not consider it further in this study.

\subsection{Other Contributions}

There are several other contributions to the EGB which we will not explicitly include in this study. For the sake of completeness, we will briefly summarize some of these possible contributions here. 

The mergers of galaxy clusters and other large scale structures can generate large-scale collisionless shocks capable of accelerating electrons to highly relativistic energies.  Through inverse Compton scattering with the cosmic microwave background, such electrons could potentially generate a non-negligible contribution to the diffuse gamma-ray background~\cite{Keshet:2002sw,Gabici:2002fg}. Assuming that $\sim$5\% of the thermal energy in such shocks is transferred to the acceleration of electrons, Ref.~\cite{Keshet:2002sw} finds that this mechanism could account for up to tens of percents of the diffuse gamma-ray background at energies above $\sim$10 GeV. In such a scenario, Fermi should be capable of detecting several merging clusters as gamma-ray sources~\cite{Keshet:2002sw,Gabici:2003kr}.  Other estimates for this contribution are significantly lower~\cite{Gabici:2002fg}, however, and it is difficult to bound the expected contribution from this mechanism. As gamma-ray emission has not yet been detected from galaxy clusters~\cite{Pfrommer:2012mm,::2013ufa,clusterfermi,Prokhorov:2013kca,Huber:2013cia}, we do not include this contribution in our model at this time.

More local phenomena could also contribute to the diffuse gamma-ray background. In particular, interactions between cosmic rays and ionized hydrogen in the outer halo of the Milky Way could produce a diffuse flux of gamma-rays capable of accounting for $\sim$1-10\% of the observed gamma-ray background~\cite{Feldmann:2012rx}. Alternatively, interactions of cosmic rays with debris in the Solar System's Oort Cloud could also contribute~\cite{Moskalenko:2009tv}. We do not include such local contributions in our calculations.

\subsection{The Combined Astrophysical Contribution to the Extragalactic Gamma-Ray Background}

\begin{figure}
\mbox{\includegraphics[width=0.49\textwidth,clip]{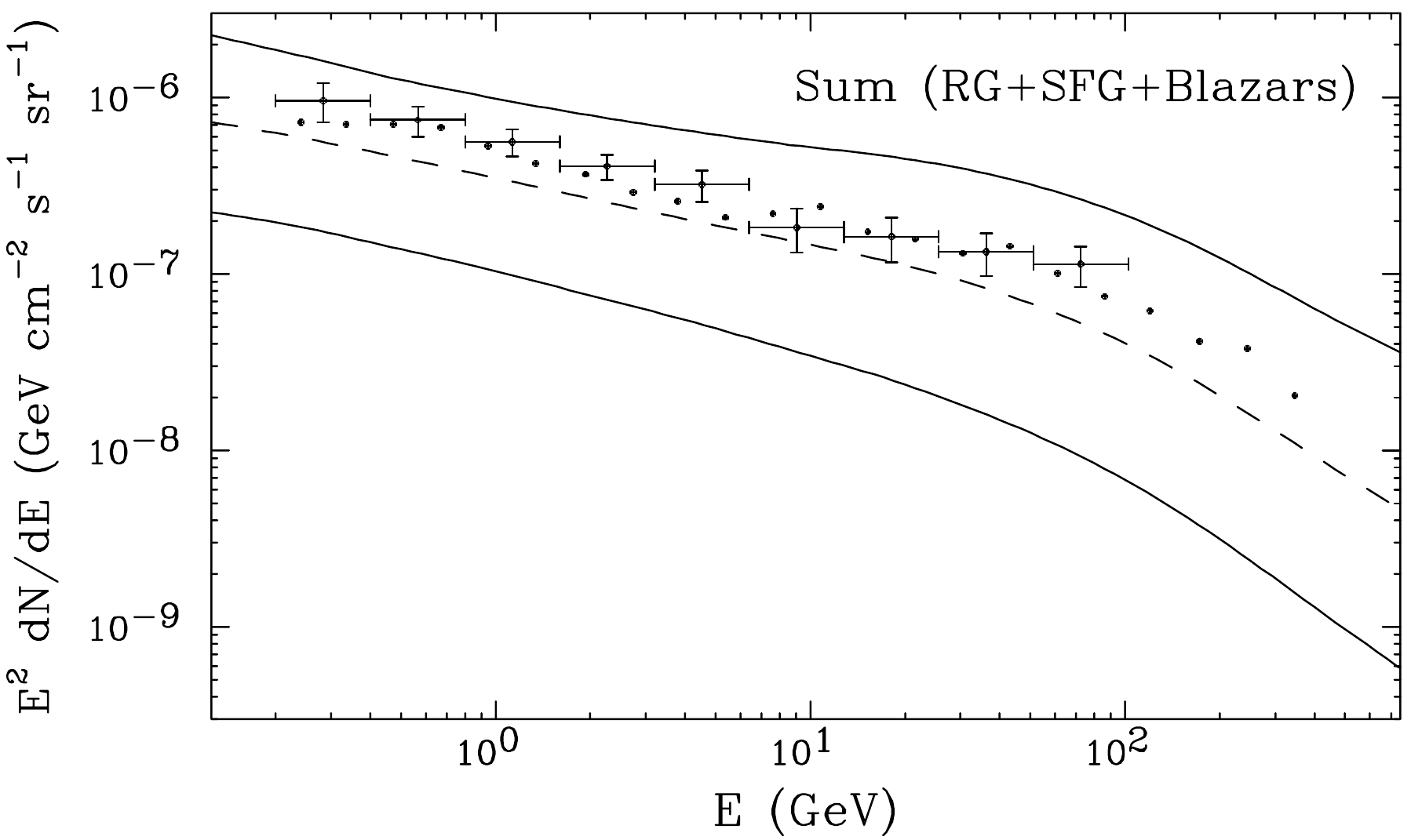}} 
\caption{The estimated contribution to the EGB from the combination of radio galaxies, star-forming galaxies, and blazars (FSRQs and Bl Lacs). The dashed contour represents the prediction using central values for all model parameters. The solid contours are the 1$\sigma$ uncertainties around this prediction, after propagating all parameter uncertainties. Error bars and points are as in Fig.~\ref{SF}. See text for details.}  
\label{sum}
\end{figure}

\begin{figure*}
\mbox{\includegraphics[width=0.49\textwidth,clip]{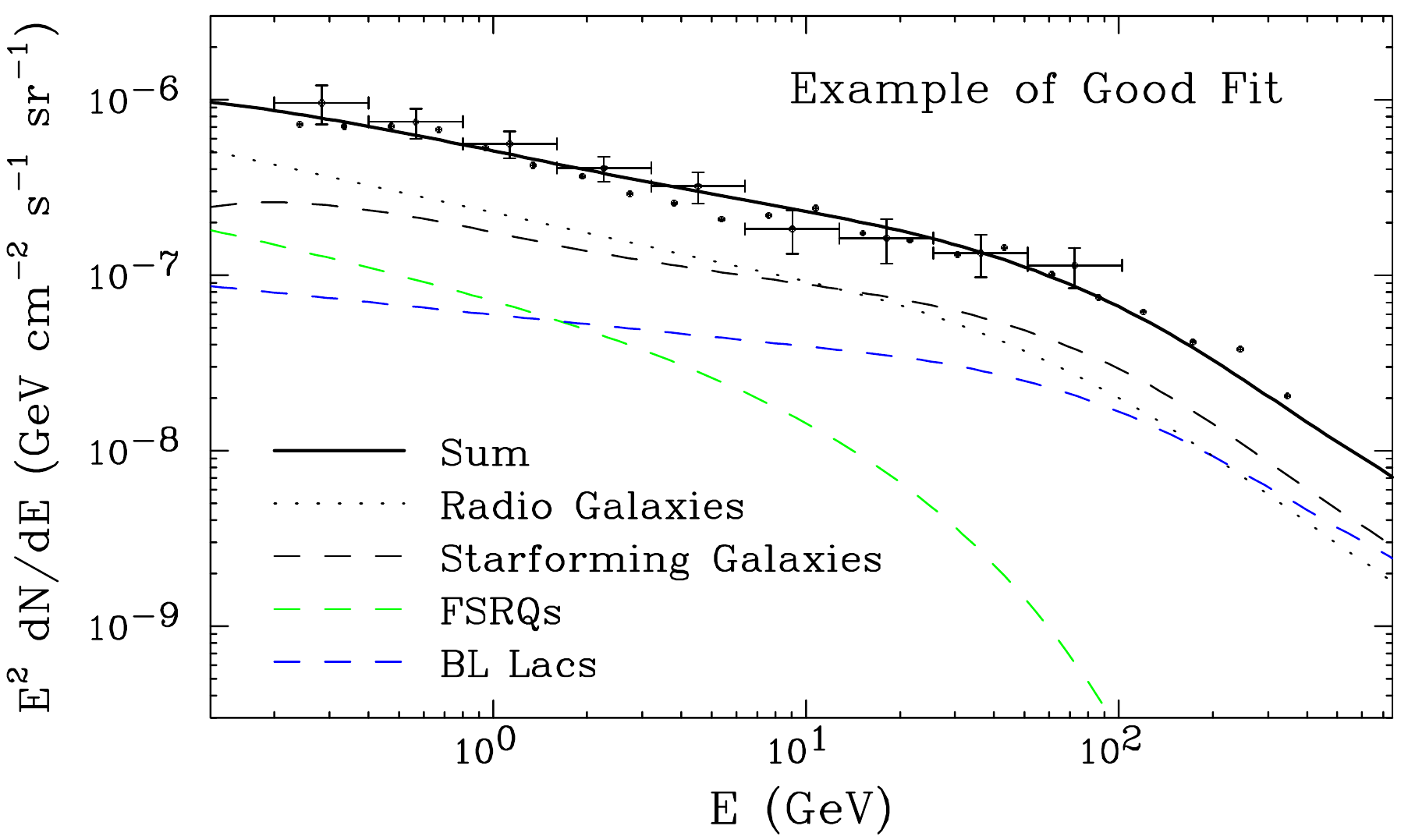}} 
\mbox{\includegraphics[width=0.49\textwidth,clip]{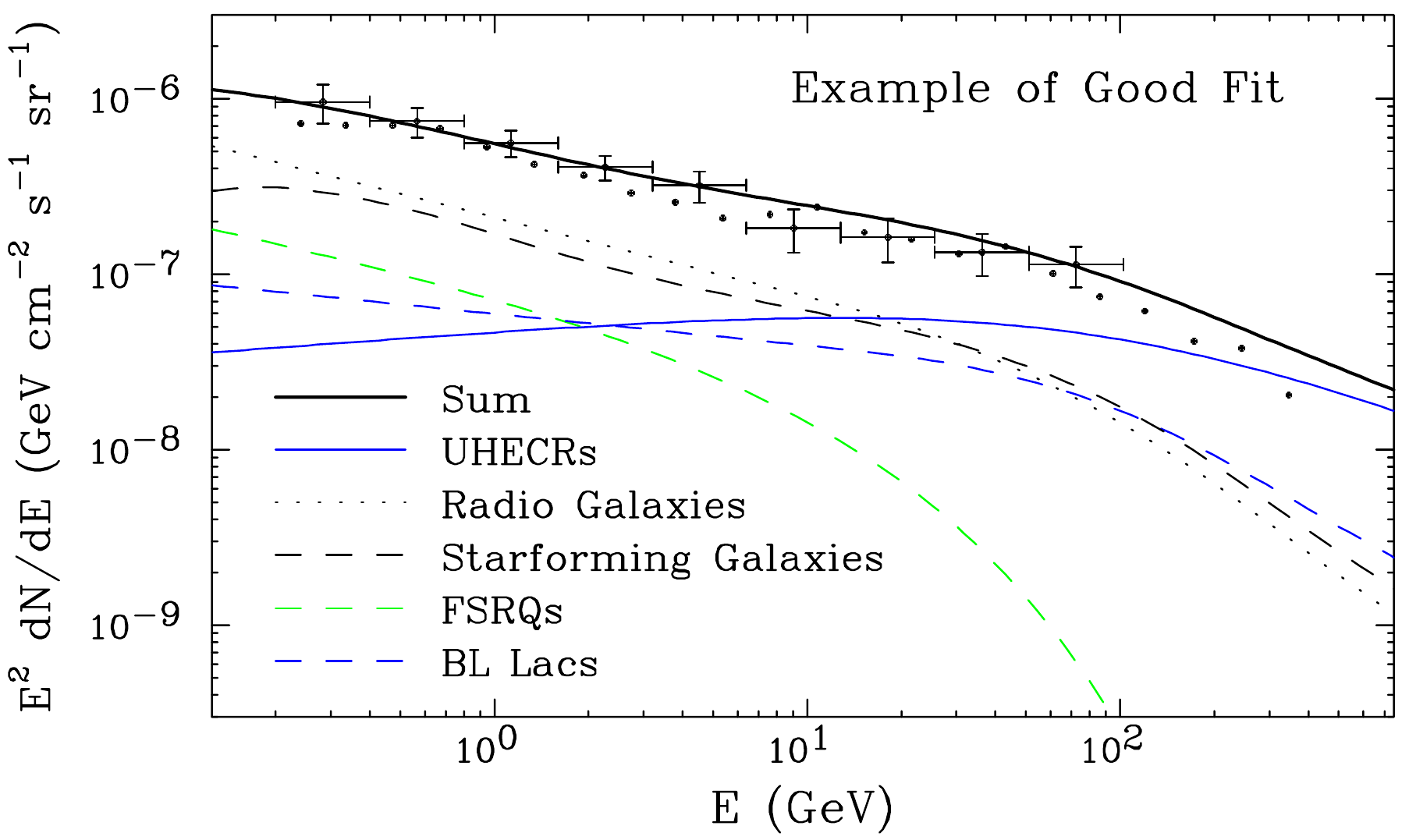}} 
\caption{Two examples of viable models which provide a good fit to the observed EGB. See text for details.}
\label{goodmodels}
\end{figure*}

In Fig.~\ref{sum}, we show the combined contributions to the EGB from radio galaxies, star-forming galaxies, FSRQs, and BL Lac objects. To evaluate a given model, we calculate the total chi-square ($\chi^2$) corresponding to all of the parameter values, as described in Secs.~\ref{sfsec}, \ref{rgsec}, and \ref{blazarsec}. The dashed curve in Fig.~\ref{sum} represents the model with central values for each parameter, whereas the solid curves denote the range covered by all models which yield a $\chi^2$ that is within $1\sigma$ of that found using the central parameter values. The result shown in Fig.~\ref{sum} does not include in its fit the spectrum of the EGB as measured by Fermi.

Remarkably, we find that the entirety of the observed EGB can be accounted for by a combination of emission from radio galaxies, star-forming galaxies, and blazars. In Fig.~\ref{goodmodels} we show two specific examples of viable astrophysical models which provide a good fit to the observed EGB. In the left frame, we show a model with a negligible contribution from UHECR propagation, whereas the model in the right frame includes a significant contribution from UHECRs (corresponding to iron nuclei primaries, with strong source evolution).  These models each yield excellent fits to the model parameters, as well as to Fermi's measurement of the EGB spectrum.

Although the astrophysical contributions included in our model are collectively able to account for the observed EGB, the uncertainties in the model remain fairly large and other contributions could also be significant. In the following sections, we calculate the isotropic gamma-ray spectrum from annihilating dark matter and include this contribution in our model of the EGB in order to derive upper limits on the corresponding annihilation cross section.

\section{Gamma Rays from Annihilating Dark Matter}
\label{dm}

If the dark matter consists of particles with weak-scale masses and cross sections, their annihilations could contribute significantly to the EGB. In this section, we examine the gamma-ray spectrum produced through dark matter particles annihilating in the halo of the Milky Way and throughout the universe. Throughout this section, we follow closely the approach of Ref.~\cite{Ando:2013ff}.

\subsection{The Extragalactic Contribution}

The intensity of the extragalactic gamma-ray background from dark matter annihilations is given by~\cite{Ando:2013ff}:
\begin{multline} \label{intens}
\frac{d^2I_{\rm eg}(E_{\gamma})}{dE_{\gamma}d\Omega} = \int  \frac{dz}{H(z)} \frac{\langle \sigma v \rangle }{8\pi m_{\rm DM}^2}  (1+z)^3 \frac{dN_\gamma}{dE_{\gamma}} e^{-\tau[E_{\gamma}(1+z),z] }  \\ \times \int dM \frac{dn(M,z)}{dM} \left[ 1+b_{\rm sh}(M,z) \right]  \int dV \rho_{\rm host}^2 (r,M,z),
\end{multline}
where $dN_\gamma/dE_{\gamma}$ is the gamma-ray spectrum per annihilation (obtained from {\tt PPPC4DMID}~\cite{Cirelli:2010xx}), $\tau$ is the optical depth (again, using the model of Ref.~\cite{Gilmore:2011ks}), $dn/dM$ is the halo mass function (which we tabulate with {\tt HMFcalc} \cite{Murray:2013qza} using the model of Ref.~\cite{Tinker:2008ff}), and $\rho_{\rm host}$ is the density profile of a given halo. As our benchmark model, we consider dark matter particles of mass $m_{\rm DM}$ and that annihilate to $b\bar{b}$. For this annihilation channel, the gamma-ray emission is dominated by the prompt photons, in contrast to contributions from inverse Compton scattering or bremsstrahlung emission, which we do not include in our calculations. In this work, we will take all host halos to have a density distribution defined by an NFW profile~\cite{NFW,NFW2}:
\beq
\rho_{\rm host} = \frac{\rho_s}{x(1+x)^2},
\eeq
where $\rho_s$ is the scale density and $x=r/r_s$ is the distance from the center of the halo in units of the scale radius, $r_s$. We relate the scale and virial radii of a halo with the concentration, $c(M,z)\equiv r_{\rm vir}/r_s$, as parameterized in Ref.~\cite{MunozCuartas:2010ig}. The mass of a halo is related to its virial radius by:
\begin{equation}
 M=\frac{4\pi}{3} r^3_{\rm vir} \Delta_{\rm vir}(z) \rho_c(z), 
\end{equation}
where $\rho_c(z)$ is the cosmological dark matter density and $\Delta_{\rm vir}(z)$ is the overdensity within the virial radius of a halo. This can be parameterized as $\Delta_{\rm vir}(z)=18 \pi^2 +82 d-39d^2$, where $d=\Omega_M (1+z)^3/[\Omega_M(1+z)^3+\Omega_{\Lambda}]-1$~\cite{Bryan:1997dn}. In terms of these quantities, the scale density is given by:
\begin{equation}
\rho_s = \frac{M}{4\pi r^3_s} \bigg[\ln(1+c)-\frac{c}{1+c} \bigg]^{-1}. 
\end{equation}
The density-squared integral in \Eq{intens} can be written as
\beq
\int dV \rho_{\rm host}^2 (r,M,z) = \frac{4\pi r_s^3\rho_s^2}3 \left[ 1-\frac1{(1+c)^3} \right],
\eeq
assuming that $\rho_{\rm host}$ is described by an NFW profile. 

The quantity $b_{\rm sh}$ accounts for the enhancement of the annihilation rate within a given halo as a result of substructures. As our default model, we consider the following parameterization for the boost factor:
\begin{equation}
b_{\rm sh}(M,z) = 110 \left(M_{200}(M,z)/10^{12}M_\odot \right)^{0.39}, 
\label{defaultboost}
\end{equation}
where $M_{200}$ is the mass of a halo contained within a region with an average density equal to 200 times the critical density (for a relationship between $M_{200}$ and $M_{\rm vir}$, see Appendix C of Ref.~\cite{Hu:2002we}).

\begin{figure}
\mbox{\includegraphics[width=0.49\textwidth,clip]{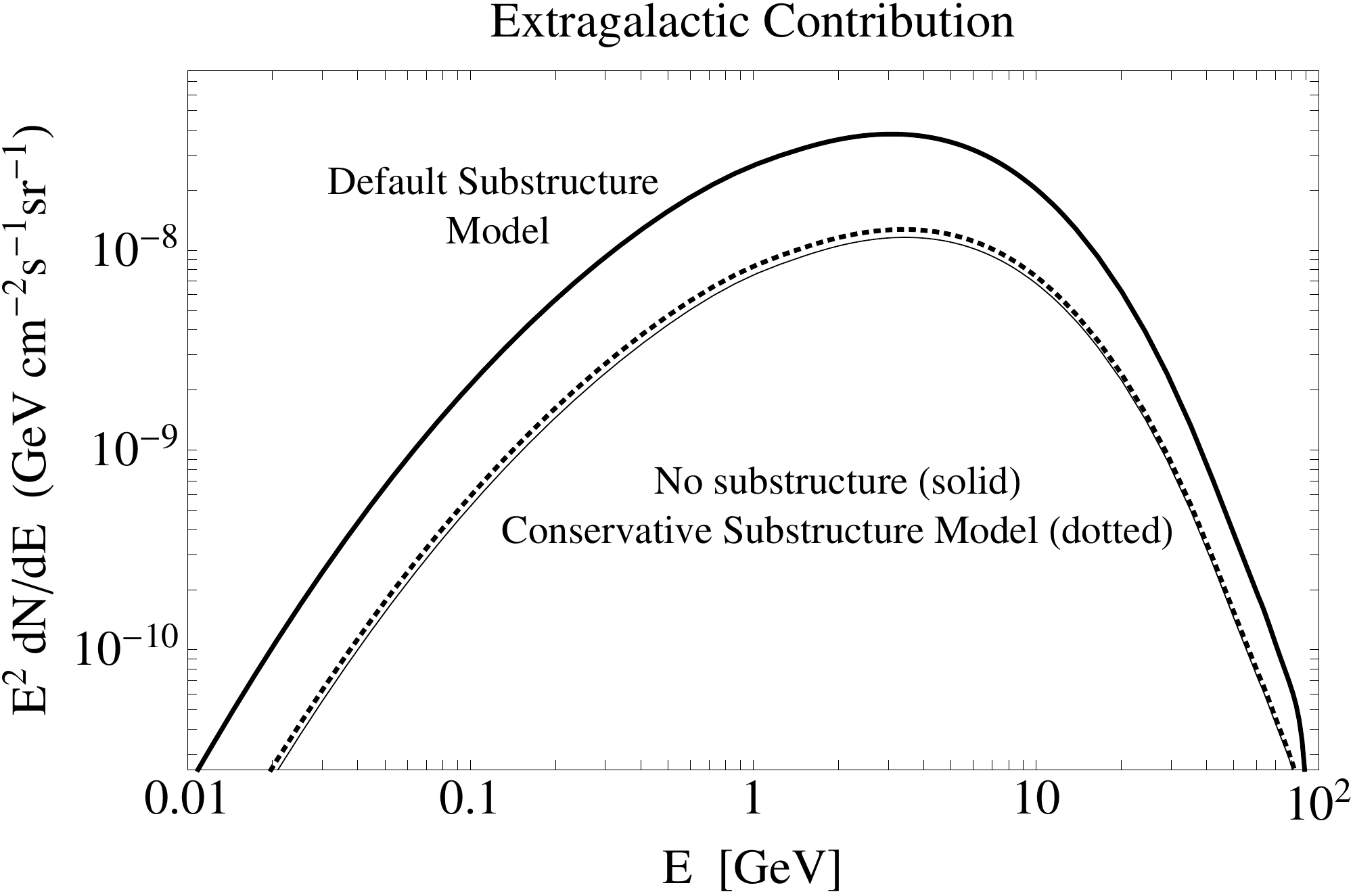}} 
\caption{The extragalactic dark matter annihilation contribution to the EGB for a reference dark matter model ($m_{\rm DM}=100$ GeV, annihilating to $b\bar{b}$ with $\sigma v=3\times 10^{-26}$ cm$^3$/s). The upper curve is the result using the substructure boost factor of Eq.~\ref{defaultboost}, which is based on an extrapolation of numerical simulations. The dotted curve assumes a boost factor that is a factor of 30 lower than our default model. The lowest curve neglects the contribution from substructure entirely. See text for details.}
\label{egal}
\end{figure}

The boost factor given in Eq.~\ref{defaultboost} was obtained from Ref.~\cite{Gao:2011rf} (modified to account for all subhalos, including those outside of the volume containing mass $M_{200}$~\cite{pcando}), and is based on the results of numerical simulations. To estimate the boost factor from such simulations, however, one must extrapolate to subhalos with masses well below the current resolution (the Aquarius simulation of Milky Way-like halos, for example, resolves subhalos with masses down to $\sim3\times 10^4\, M_{\odot}$~\cite{Springel:2008cc}).  In particular, the result of Eq.~\ref{defaultboost} assumes that the subhalo mass function extends down to a minimum mass of $M_{\rm min}=10^{-6}\, M_{\odot}$, and that the mass-concentration relationship observed among very massive simulated subhalos can be extrapolated to much smaller subhalos. In regards to the minimum subhalo mass, the precise value of $M_{\rm min}$ is determined by the temperature at which the dark matter particles decouple kinetically from the cosmic neutrino background. And while the value of $M_{\rm min}$ is model-dependent, typical dark matter candidates with masses and annihilation cross sections in the range of interest to this study generically yield minimum masses in the range of $M_{\rm min} \sim 10^{-9} - 10^{-3} \, M_\odot$~\cite{Profumo:2006bv,Cornell:2013rza}.  If we had increased the minimum subhalo mass assumed from $10^{-6}$ to $10^{-3}$ solar masses, for example, the boost factors would be reduced by a factor of $\sim$4 relative to those given by Eq.~\ref{defaultboost}. Of potentially greater importance, however, is the extrapolation of the subhalo mass-concentration relationship. If the concentrations of low mass subhalos are not as large as suggested by current extrapolations, the resulting boost factors could be very significantly reduced. As an example of the variation found in the literature, we note that the boost factors presented in Ref.~\cite{Anderhalden:2013wd} for galaxy-sized halos are a factor of $\sim$30 smaller than those described in Eq.~\ref{defaultboost}. With this in mind, we plot in Fig.~\ref{egal} the contribution to the EGB from extragalactic dark matter annihilations, for a reference dark matter model ($m_{\rm DM}=100$ GeV, annihilating to $b\bar{b}$ with $\sigma v=3\times 10^{-26}$ cm$^3$/s), and for three sets of assumptions regarding substructure. The upper curve is our default case (Eq.~\ref{defaultboost}), while the lower dotted curve represents a more conservative case in which the boost factor is reduced by a factor of 30. We also show a calculation which entirely neglects the contribution from substructure; this is shown as the lower solid line in \Fig{egal}. We note that the conservative case is almost indistinguishable from the case in which we neglect substructures entirely.

\begin{figure}
\mbox{\includegraphics[width=0.49\textwidth,clip]{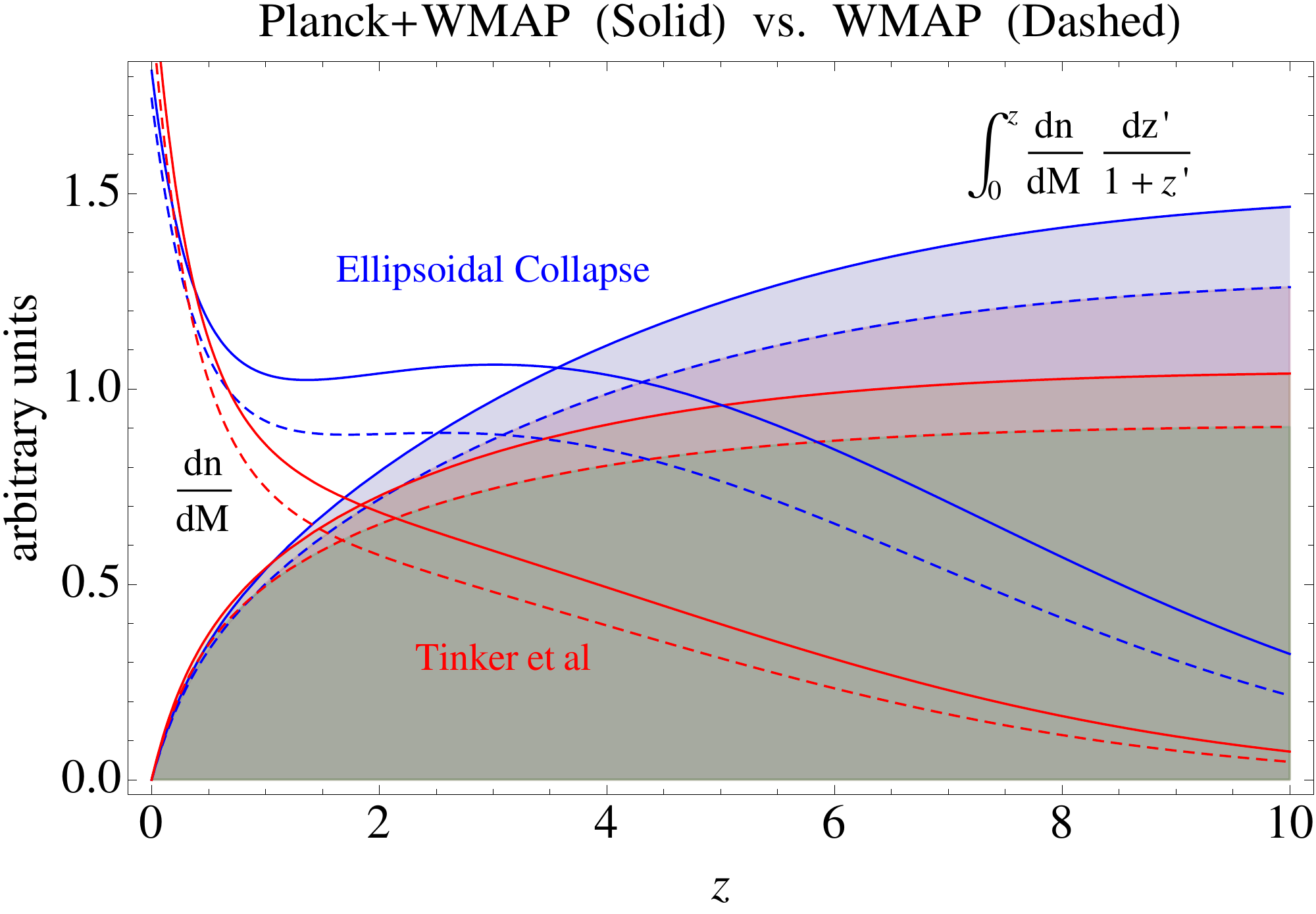}} 
\caption{The halo mass function, $dn/dM$, and the integral of the $(1+z)^{-1}$ weighted halo mass function using the model of Tinker et al.~\cite{Tinker:2008ff} (adopted in our calculations) and the ellipsoidal collapse model adopted in Ref.~\cite{Ando:2013ff}. We also show results using pre-Planck (dashed) and post-Planck (solid) values for the relevant cosmological parameters. These differences have only a modest impact on the contribution of dark matter annihilations to the extragalactic gamma-ray background.}
\label{changes}
\end{figure}

We briefly mention that our results are slightly different from those of Ref.~\cite{Ando:2013ff}, due to differences in our underlying assumptions. Firstly, the authors of Ref.~\cite{Ando:2013ff} adopted a halo mass function based on an ellipsoidal collapse model, whereas we have instead adopted the model of Ref.~\cite{Tinker:2008ff}. Secondly, we have updated our cosmological parameters to include the recent results of the Planck experiment~\cite{Ade:2013zuv}. In Fig.~\ref{changes}, we show that the combined impact of these differences changes the overall normalization of the extragalactic dark matter signal by a factor of less than $\sim$20\% relative to the results of Ref.~\cite{Ando:2013ff}.

\subsection{The Smooth Galactic Halo}

The angle-averaged intensity from dark matter annihilations in the halo of the Milky Way (neglecting substructures) is given by:
\beq \label{smoothfinal}
\left< \frac{dI_{\rm sm}(E_{\gamma})}{dE_{\gamma}} \right>= \frac{\langle \sigma v \rangle}{2 m_{\rm DM}^2} \frac{dN_\gamma}{dE_{\gamma}}  \frac1{\Omega_e}  \int_{V_*} dV ~ \frac{ \rho^2(s,b,\ell)}{4\pi s^2},
\eeq
where $s$ is the distance from the center of the halo, $b$ and $\ell$ are the direction in galactic coordinates, and $\Omega_e$ is the solid angle observed. We take the dark matter to be distributed according to an NFW profile, and we adopt parameters consistent with measurements: $r_s=21.5$ kpc, $r_{\rm vir}=258$ kpc, and $M_{\rm vir}=1.0\times 10^{12}\, M_{\odot}$~\cite{Klypin:2001xu}. These parameters imply a local dark matter density of $\rho_{\odot} \approx 0.24$ GeV cm$^{-3}$, which is somewhat low compared to the more canonical estimates of 0.3-0.4 GeV/cm$^3$~\cite{Iocco:2011jz, Bovy:2012tw,Catena:2009mf, Salucci:2010qr}. If we had scaled up the dark matter density to a value in this range, the local annihilation rate would be further enhanced by a factor of $\sim$1.6-2.8. 

The galactocentric radius is related to the distance along the line-of-sight by
\beq
r^2 = s^2 + r_\odot^2 - 2 s r_\odot \cos b \cos \ell.
\eeq
The solid angle of interest is described by $0\leq \ell <2\pi$ and $30^\circ\leq |b| \leq 90^{\circ}$.

\begin{figure}
\mbox{\includegraphics[width=0.49\textwidth,clip]{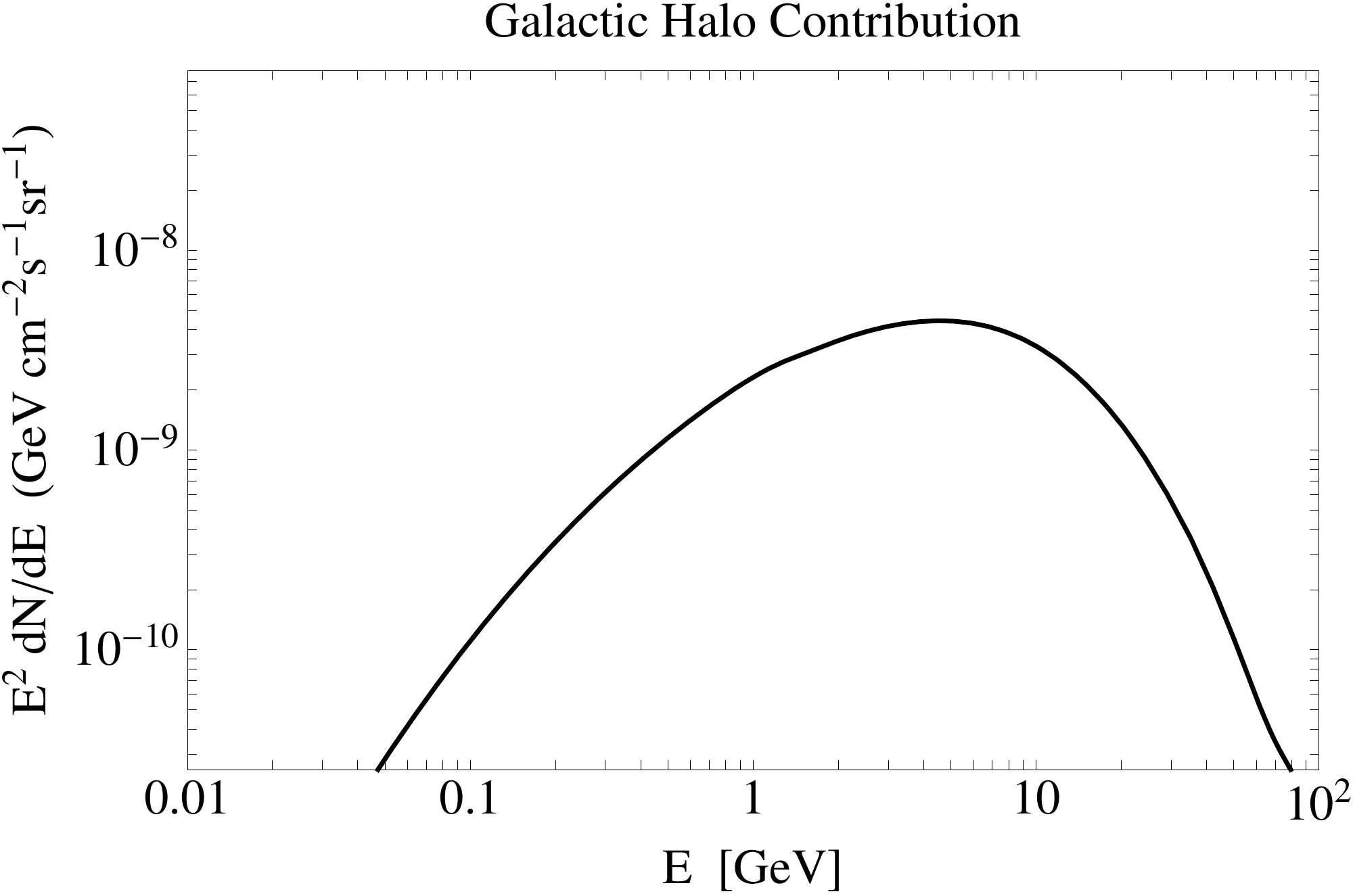}} 
\caption{The contribution to the extragalactic gamma-ray background from dark matter annihilations in the smooth halo of the Milky Way, for a reference dark matter model ($m_{\rm DM}=100$ GeV, annihilating to $b\bar{b}$ with $\sigma v=3\times 10^{-26}$ cm$^3$/s). The result has been averaged over the following region of the sky: $0< \ell < 2 \pi$ and $|b| > 30^\circ$.  See text for details.}
\label{mwsm}
\end{figure}

In Fig.~\ref{mwsm}, we plot the contribution to the EGB from dark matter annihilations in the smooth component of the Milky Way's halo.  Comparing this to the extragalactic contribution, we find that this component is likely to be subdominant, even for conservative assumptions pertaining to extragalactic substructure.

\subsection{Subhalos of the Milky Way}

Although the smooth halo of the Milky Way is predicted to provide no more than a subdominant contribution to the EGB, the intensity of gamma rays from dark matter annihilations in the subhalos of the Milky Way are expected to be comparable to the intensity of gamma rays from extragalactic structures. Each subhalo has a differential luminosity which is totally determined by its density profile:
\beq \label{sat:lum}
\frac{dL_{\gamma}}{dE_{\gamma}}=\frac{\langle \sigma v \rangle}{2 m_{\rm DM}^2} \frac{dN_\gamma}{dE_{\gamma}} \int dV\, \rho_{\rm sub}^2.
\eeq
 For a subhalo of mass, $M$, at a distance, $s$, along the line-of-sight, the photon intensity at earth is given by:
\begin{eqnarray} \label{sat:individual}
\frac{di(E_{\gamma},s,M)}{dE_{\gamma}}&=& \frac1{4\pi s^2} \frac{d L(E_{\gamma},\langle \sigma v \rangle,m_{\rm DM},M)}{dE_{\gamma}} \\
&=&\frac1{4\pi s^2} \frac{b_{\rm gs} \langle \sigma v \rangle}{2 m_{\rm DM}^2} \frac{dN_\gamma}{dE_{\gamma}}  \frac{M^2}{r_s(M)^3} g[c(M)] , \nonumber
\end{eqnarray}
where $r_s$ is the scale radius of the subhalo and $b_{\rm gs}$ describes the contribution from substructure within each subhalo, which we set equal to 2, irrespective of mass \cite{Kuhlen:2008aw}. The function $g[c(M)]$ arises from the integral over the volume of each satellite. For our default calculation, we set the subhalo concentrations following the approach of Ref.~\cite{Ando:2009fp}, where the subhalo is assumed to be initially described by an NFW profile which is then tidally stripped, leaving only a very compact and dark matter-dominated object. In this case,
\beq
g[c(M)]=\frac1{12\pi} \bL 1 - \frac1{\pL 1 + c\pR^3} \bR \bL \ln(1+c)-\frac {c}{1+c} \bR^{-2},
\eeq
where $c$ is the concentration of the subhalo. In addition to our default assumptions, we also consider a more conservative scenario in which the contribution from galactic subhalos is suppressed by a factor of 30 relative to our default case, motivated by analogy to the extragalactic calculation.

The total intensity of gamma rays at Earth from dark matter particles annihilating in galactic subhalos is then given by integrating Eq.~\ref{sat:individual} over the distribution of Milky Way subhalos. Thus we have
\begin{equation}
\frac{dI_{\rm sub}(E_{\gamma})}{dE_{\gamma}}=\int dVdM \frac{dn_{\rm sub}(M,s,\ell,b)}{dM}\frac{di(E_{\gamma},s,M)}{dE_{\gamma}},
\end{equation}
where $\int dMdV (dn_{\rm sub}/dM)$ is the total number of subhalos in the Milky Way. We assume that the subhalo mass function, $dn_{\rm sub}/dM$, is given by the anti-biased case of Ref.~\cite{Ando:2009fp}, which is proportional to an Einasto profile with $\alpha_E = 0.68$.

\begin{figure}
\mbox{\includegraphics[width=0.49\textwidth,clip]{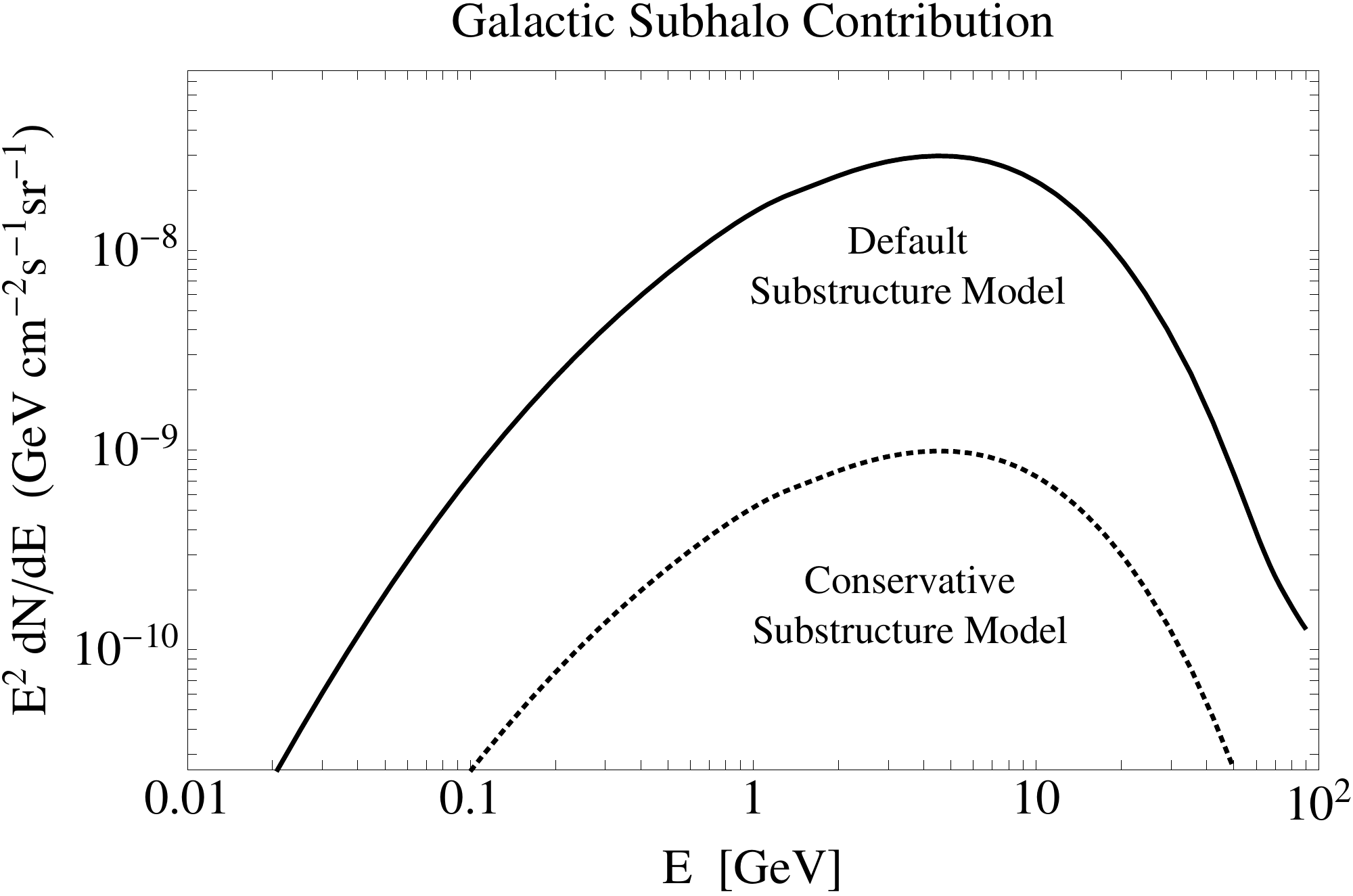}} 
\caption{The contribution to the EGB from subhalos of the Milky Way, for a reference dark matter model ($m_{\rm DM}=100$ GeV, annihilating to $b\bar{b}$ with $\sigma v=3\times 10^{-26}$ cm$^3$/s). The upper curve is the result using our default model, while the lower dotted curve is reduced by a factor of 30 relative to our default model. See text for details.}
\label{mwsh}
\end{figure}

To compare to observations, we are interested in the angle-averaged intensity of gamma rays per unit energy over the entire galaxy. This is given by:
\begin{eqnarray}
\label{first}
\left< \frac{dI_{\rm sub}(E_{\gamma})}{dE_{\gamma}} \right> &=&  \frac1{\Omega_e}  \int_{M_*} \int_{V_*(M)} dVdM \times \\ \nonumber
 && \times \frac{dn_{\rm sub}(M,s,\ell,b)}{dM}\frac{di(E_{\gamma},s,M)}{dE_{\gamma}},
\end{eqnarray}
where $V_*$ is the volume beyond which satellites are not resolved. We consider subhalos with masses in the range of $10^{-6} M_\odot \leq M_* \leq  10^{10} M_\odot$, and assume that they are not resolvable beyond a distance of $s_*(M)=\sqrt{L(M)/4\pi F_{\rm sens}}$, where $F_{\rm sens}=2\times 10^{-10} \cm^{-2} \sec^{-1}$ \cite{Ando:2009fp} and $L(M)$ is the integral of  \Eq{sat:lum} over all energy.

\begin{figure}
\mbox{\includegraphics[width=0.49\textwidth,clip]{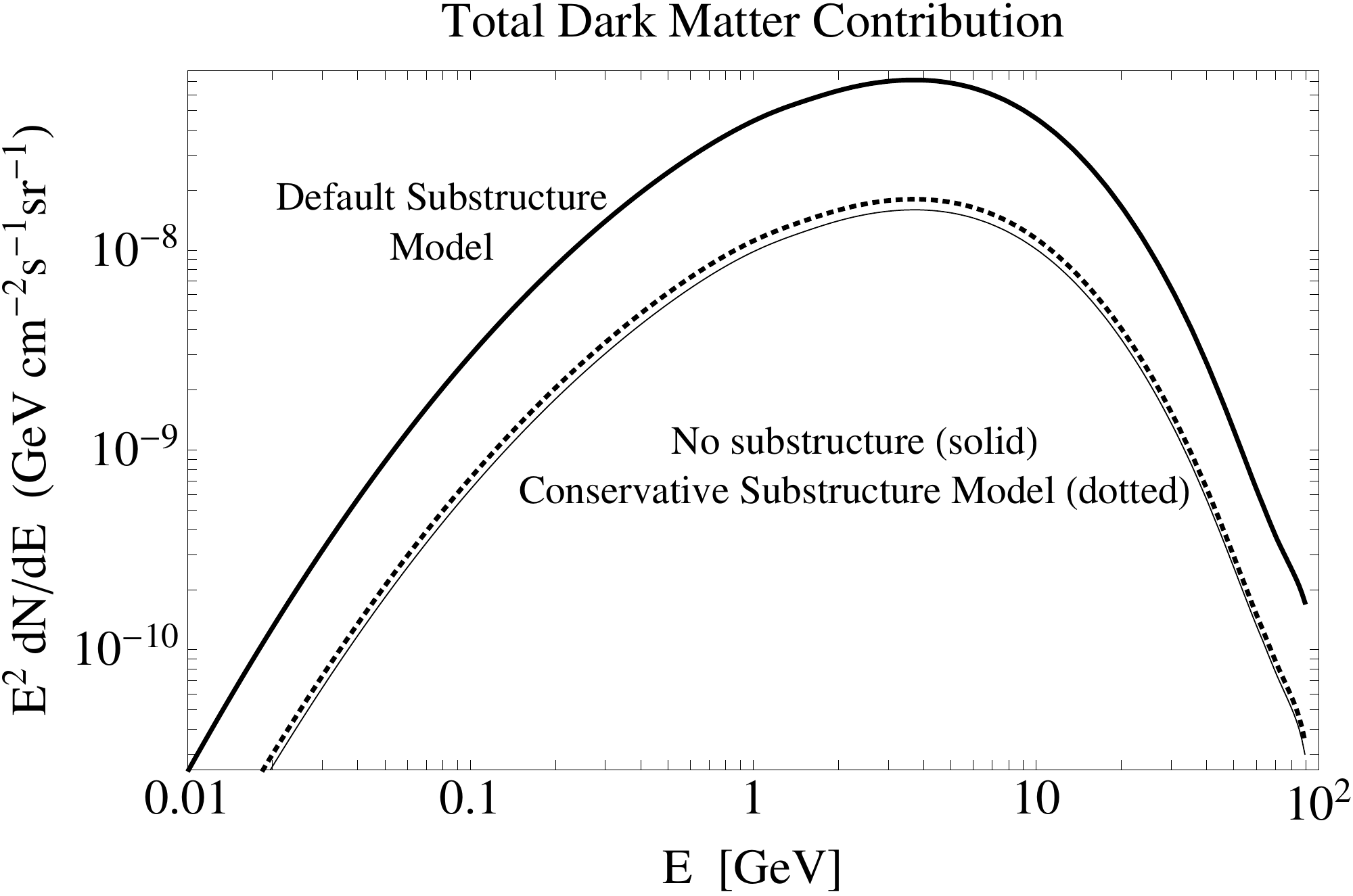}} 
\caption{The total contribution from dark matter annihilations to the EGB, for a reference dark matter model ($m_{\rm DM}=100$ GeV, annihilating to $b\bar{b}$ with $\sigma v=3\times 10^{-26}$ cm$^3$/s). The upper curve is the result using our default substructure model, while the lower dotted curve is reduces the contribution from substructure by a factor of 30 relative to our default model. See text for details.}
\label{DMtot}
\end{figure}

In Fig.~\ref{mwsh}, we show the contribution to the EGB from galactic subhalos. For our default substructure model, this contribution is comparable to that from extragalactic dark matter annihilations. In our conservative substructure model, galactic subhalos are negligible compared to the EGB. 

A summary of this section's results is given in Fig.~\ref{DMtot}. Here, we have plotted the combination of extragalactic, smooth galactic, and galactic subhalo contributions to the EGB. The upper solid curve adopts our default substructure model. The lower dotted and solid curves use our conservative substructure model or neglect substructure entirely, respectively. We note that contributions to the EGB from subhalos in the Milky Way and from extragalactic structure can be reduced significantly if low-mass halos and subhalos are not as highly concentrated as is suggested by extrapolations of simulations.  The contribution from the smooth halo of the Milky Way, however, is significantly more robust. We also remind the reader that we have conservatively adopted a relatively low value of density of dark matter in the Milky Way (corresponding to a local density of 0.24 GeV cm$^{-3}$). The more conservative models reduce the overall gamma ray flux from dark matter annihilations by only a factor of $\sim$4 relative to our default model.

\begin{figure*}
\mbox{\includegraphics[width=0.49\textwidth,clip]{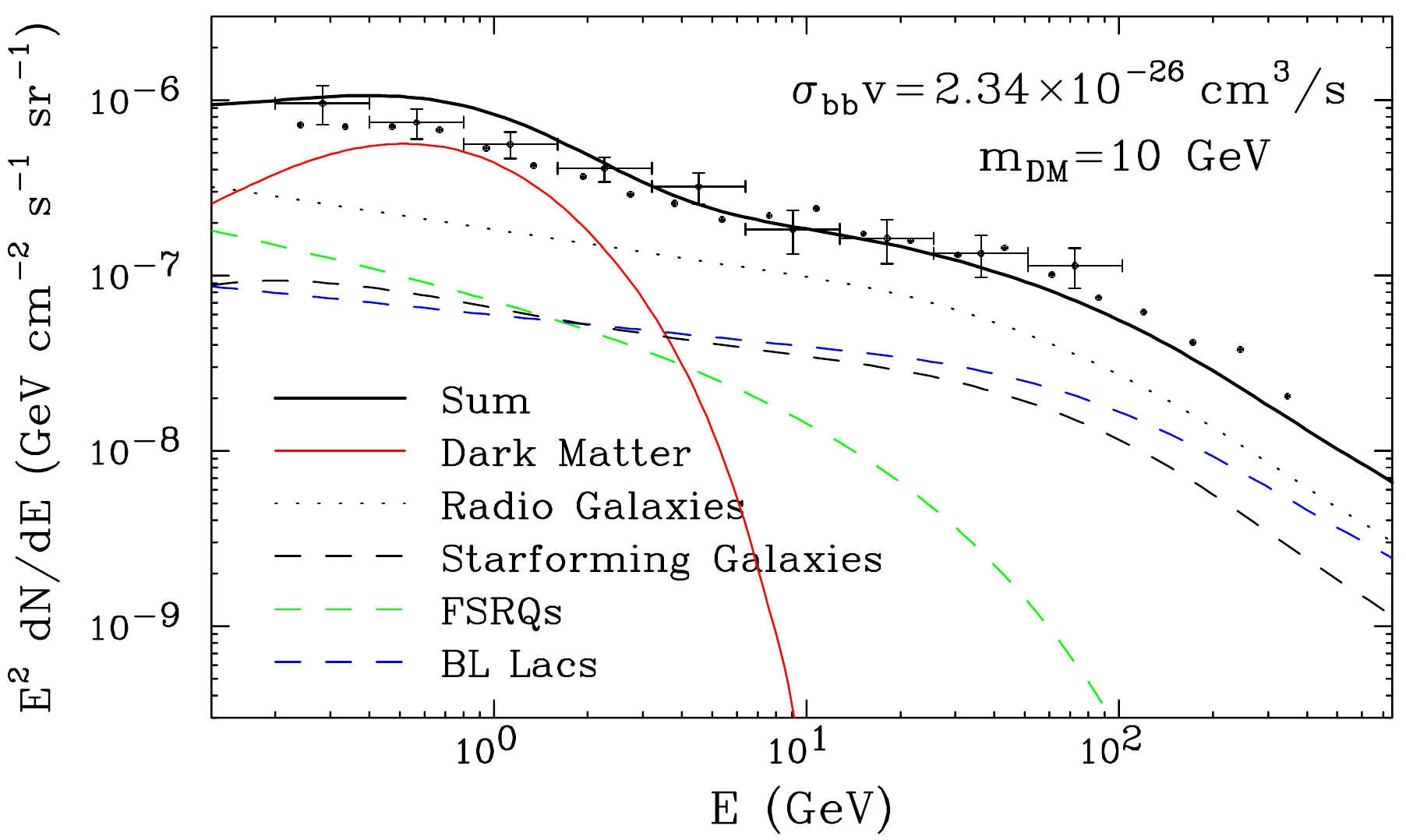}} 
\mbox{\includegraphics[width=0.49\textwidth,clip]{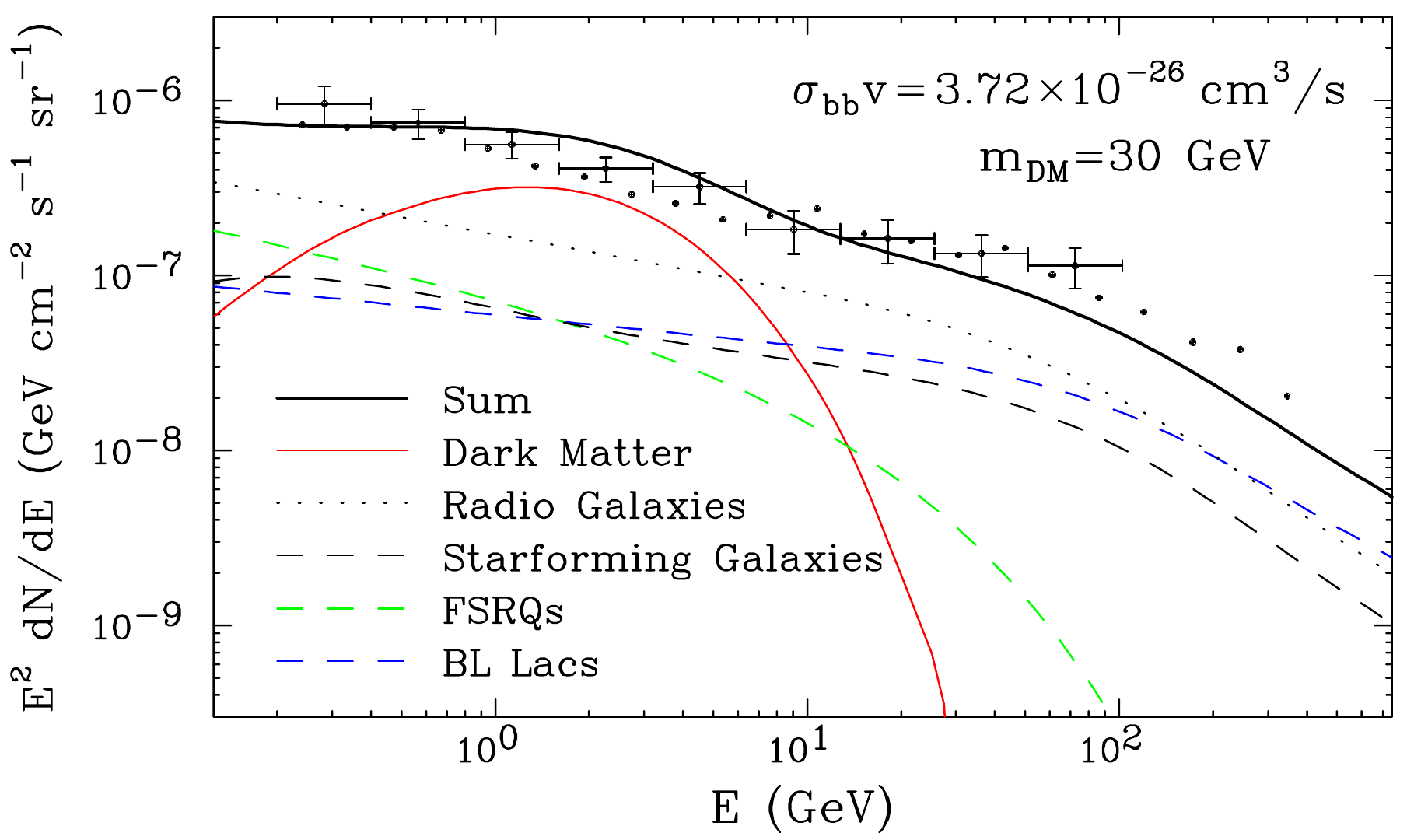}} \\
\mbox{\includegraphics[width=0.49\textwidth,clip]{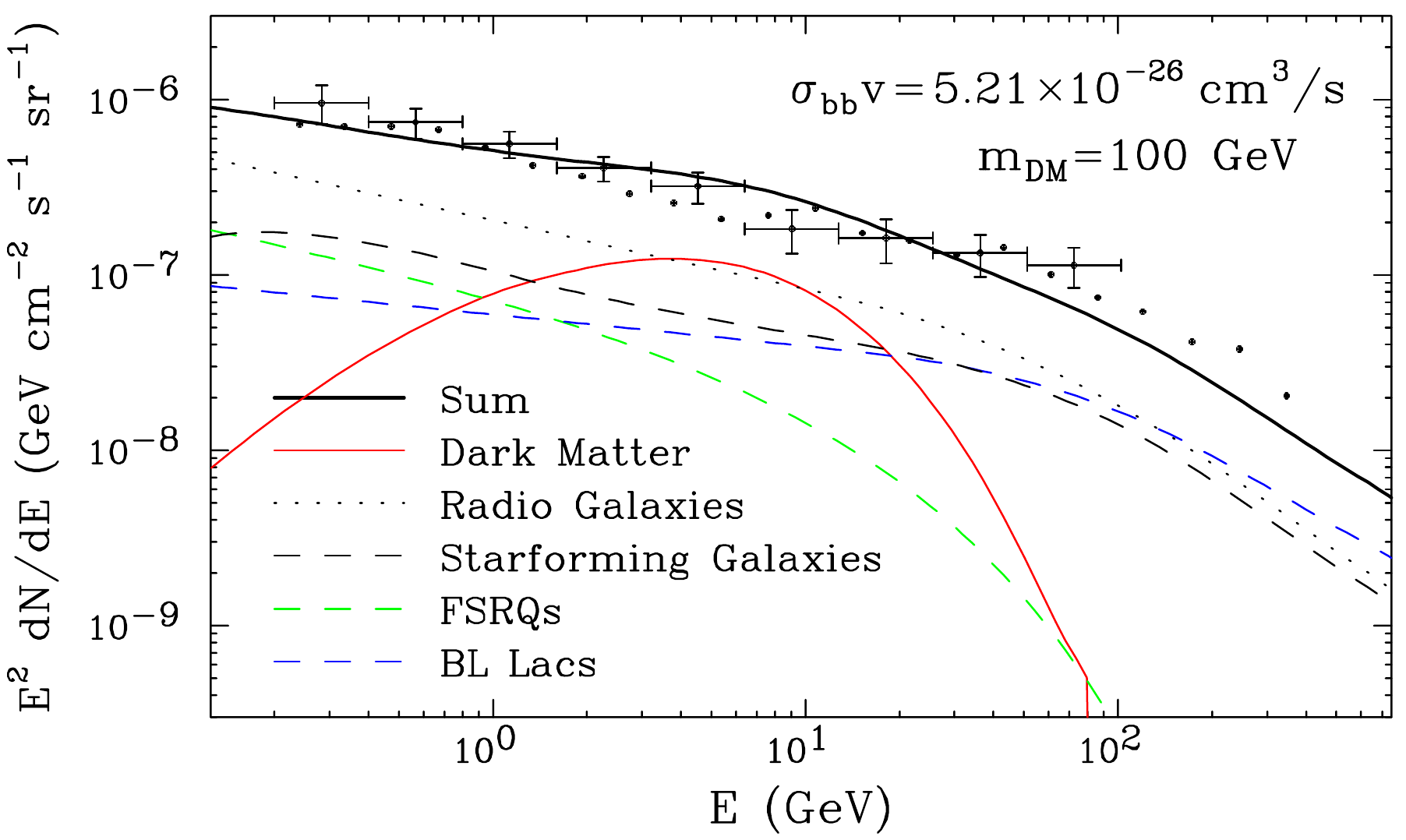}} 
\mbox{\includegraphics[width=0.49\textwidth,clip]{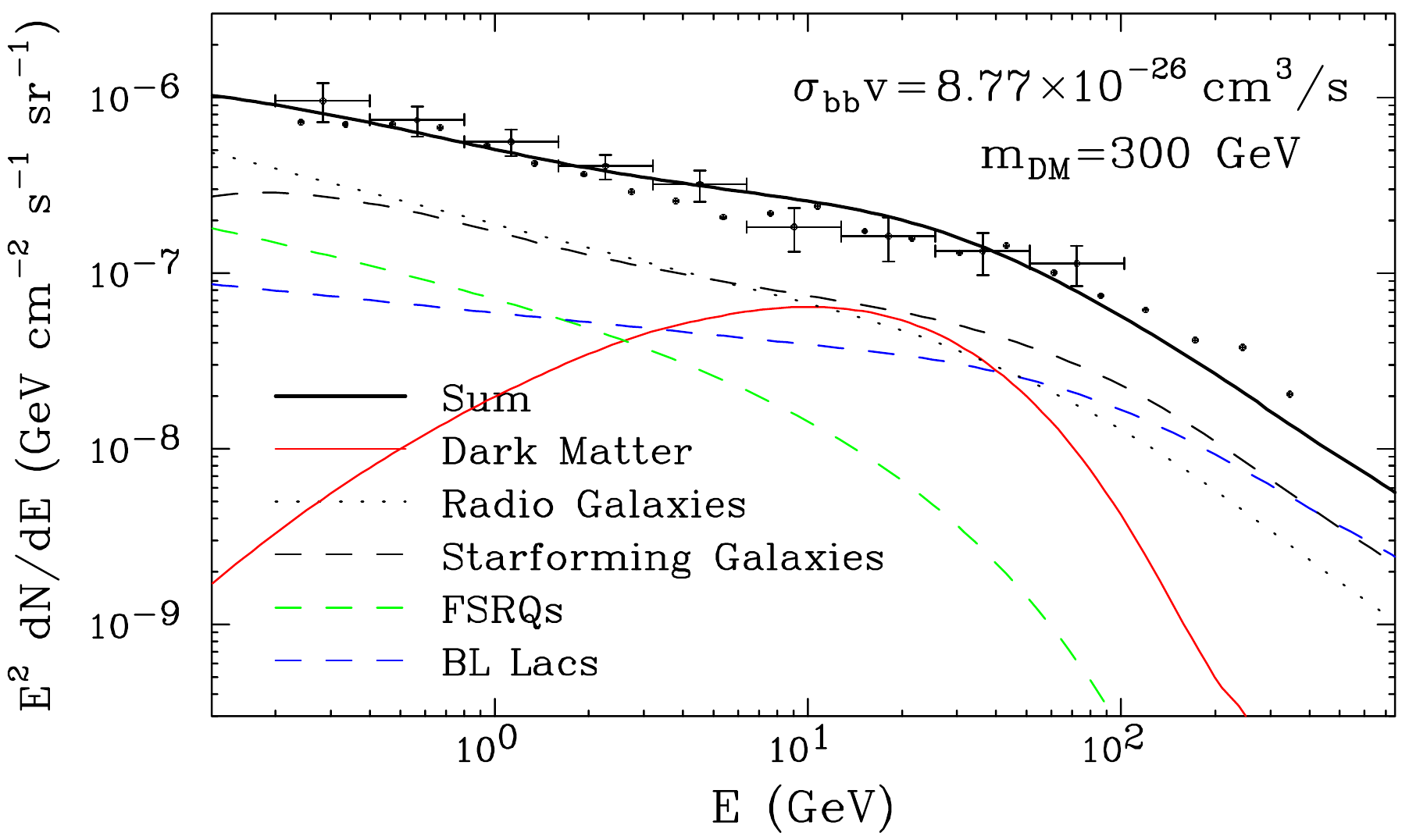}} \\
\mbox{\includegraphics[width=0.49\textwidth,clip]{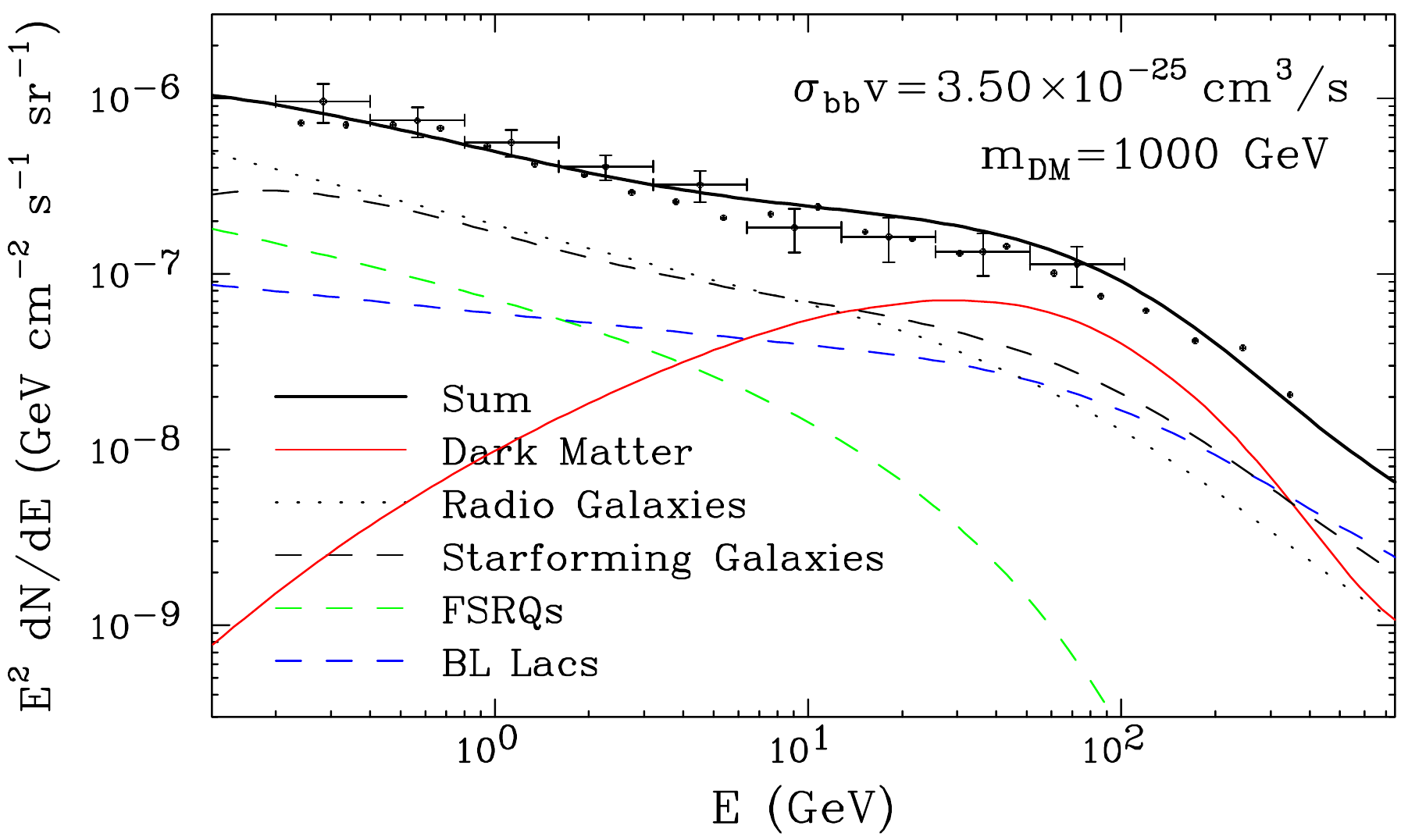}} 
\caption{Our model of the EGB, including the largest allowed contribution from annihilating dark matter (at the 95\% CL). Here, we have adopted our default substructure model.  In each case, we have marginalized over the parameters of our astrophysical model. See text for details.}
\label{fit}
\end{figure*}

\section{Constraints on the Dark Matter Annihilation Cross Section}
\label{constraints}

\begin{figure*}
\mbox{\includegraphics[width=0.49\textwidth,clip]{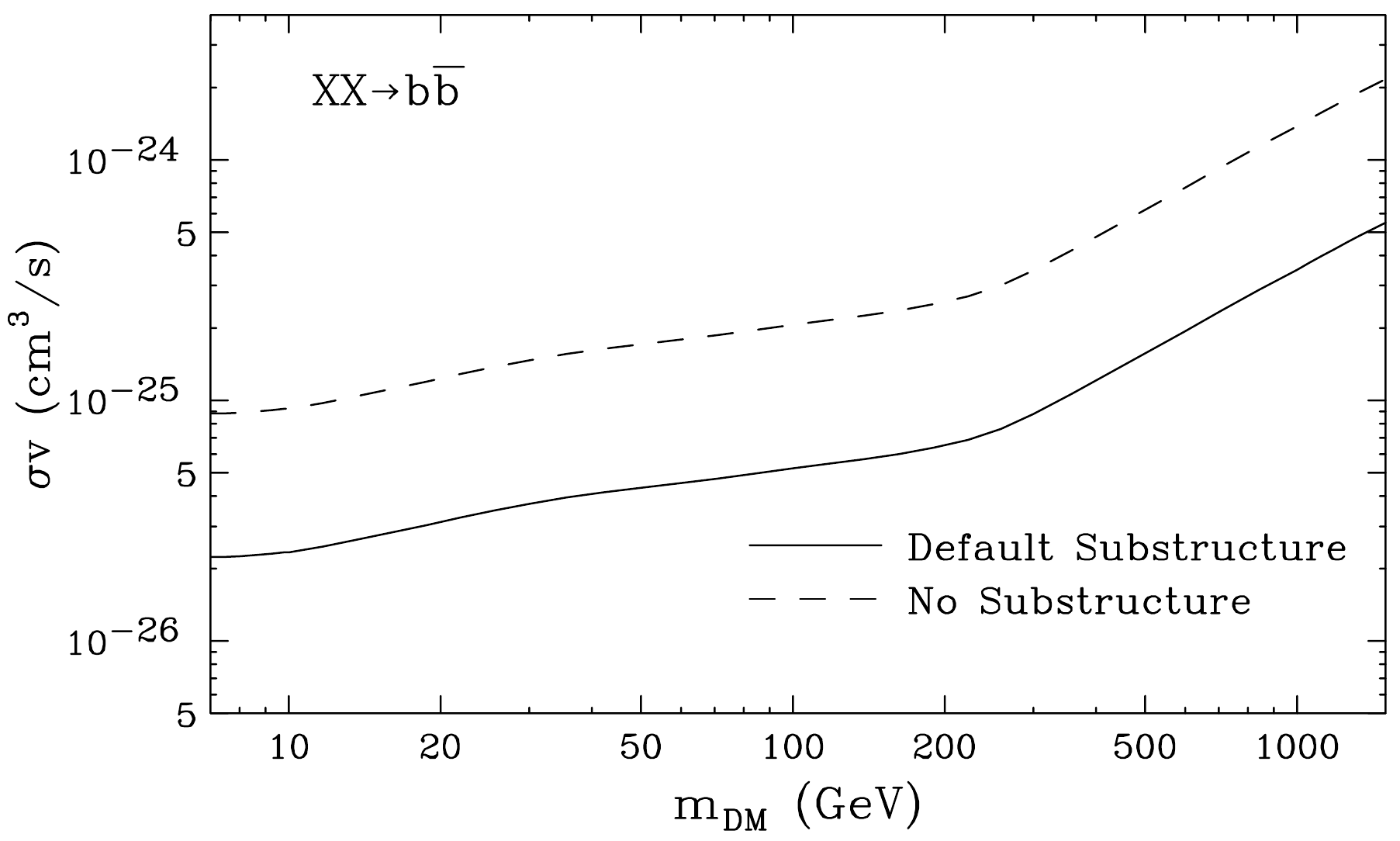}} 
\mbox{\includegraphics[width=0.49\textwidth,clip]{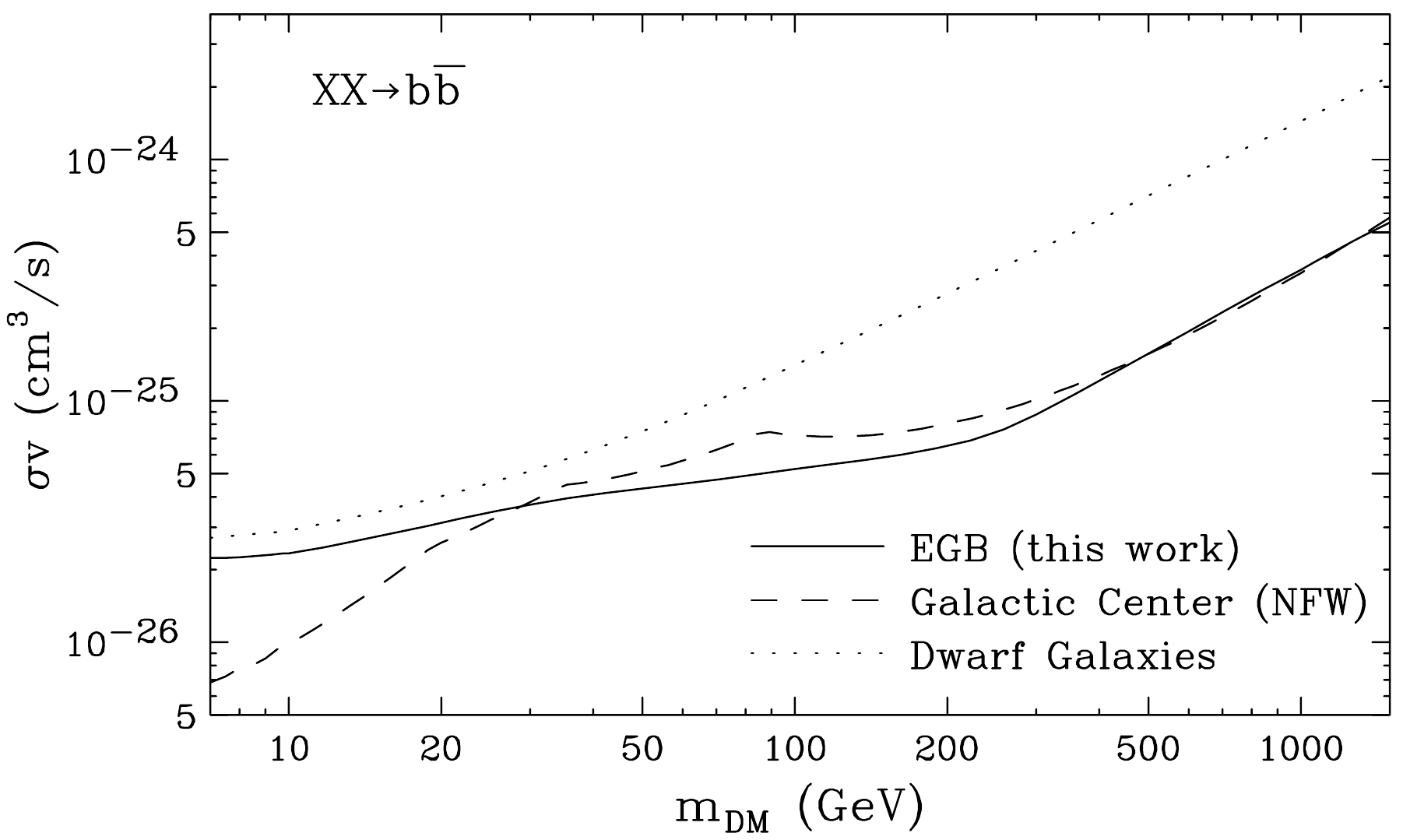}} 
\caption{In the left frame, we show the limits (95\% CL) on the dark matter annihilation cross section derived in this study, using our default substructure model (solid), and neglecting substructure (dashes). In the right frame, we compare this result to the strongest existing constraints on the dark matter annihilation cross section from observations of the Galactic Center~\cite{Hooper:2012sr} and of dwarf spheroidal galaxies~\cite{Ackermann:2013yva}. See text for details.}
\label{limits}
\end{figure*}

In this section, we combine the results of Secs.~\ref{astro} and~\ref{dm} in order to place constraints on the contribution from annihilating dark matter to the EGB (for previous dark matter constraints derived from the EGB, see Refs.~\cite{Ando:2013ff,Abazajian:2010zb,Calore:2013yia}). 

We begin by assessing the ability of a given model to fit the observed data. To do this, we construct a $\chi^2$ statistic:
\begin{equation}
\chi^2= \sum_i \frac{(p_i - p_{i,0})^2}{\sigma^2_{p,i}} + \sum_j \frac{(d_j -d_{j,0})^2}{\sigma^2_{d,j}},
\end{equation}
where the first sum is performed over the astrophysical parameters of the model ($p_i$), as described in Sec.~\ref{astro}, and the second sum is performed over the the error bars of the EGB spectrum as reported by the Fermi collaboration~\cite{Abdo:2010nz}. The quantities $\sigma_{p,i}$ and $\sigma_{d,j}$ represent the uncertainties in the astrophysical parameters and the errors in the measured spectrum, respectively. With no contribution from dark matter, our best model parameter set yields an overall value of $\chi^2=8.54$. This model includes contributions from radio galaxies, star-forming galaxies, FSRQs, and BL Lac objects, with uncertainties in the model parameters as described in Sec.~\ref{astro}.

To place limits on the dark matter annihilation cross section, we add a contribution from annihilations of dark matter (with a given mass and annihilation channel) to our model. We increase the value of the cross section until the best possible $\chi^2$ (marginalizing over all the parameters of the astrophysics model) increases by 2.71 over the best-fit with no dark matter component (corresponding to the 95\% confidence level upper limits). In Fig.~\ref{fit}, we show the contributions to the EGB in models with the maximum allowed contribution from annihilating dark matter (assuming annihilations to exclusively to $b\bar{b}$ for five choices of the dark matter mass).

In the left frame of Fig.~\ref{limits}, we plot the upper limits on the dark matter annihilation cross section  derived in this study. In the right frame, this result is compared to the limits obtained from observations of the Galactic Center~\cite{Hooper:2012sr} and of dwarf spheroidal galaxies~\cite{Ackermann:2013yva}. For our default substructure model, the limits presented here are approximately as stringent as those derived from the Galactic Center (assuming an NFW profile). Our limits obtained neglecting contributions from substructure are comparably stringent to those derived from the Galactic Center assuming a profile with a kiloparsec-scale core~\cite{Hooper:2012sr}. And although the constraint from dwarf galaxies is somewhat less susceptible to astrophysical uncertainties than those derived from the EGB or Galactic Center, even for very conservative assumptions ({\it i.e.}. negligible contributions from substructure) the constraints derived here are as or more sensitive to dark matter particles with masses on the order of 100 GeV or greater.

\section{Projections And Future Sensitivity}
\label{projections}

As Fermi continues to collect data, its sensitivity to dark matter annihilation products in the EGB will increase due to two different sets of factors. Firstly, Fermi's measurement of the EGB itself will improve, reducing the errors on the corresponding spectrum and extending the measurement to higher energies. Secondly, with a larger data set, Fermi will detect GeV emission from a greater number of radio galaxies, star-forming galaxies, and blazars, and will characterize the emission from those sources already detected with greater precision. As it does so, the uncertainties in the contributions to the EGB from these sources classes will be reduced considerably.

\begin{figure*}
\mbox{\includegraphics[width=0.49\textwidth,clip]{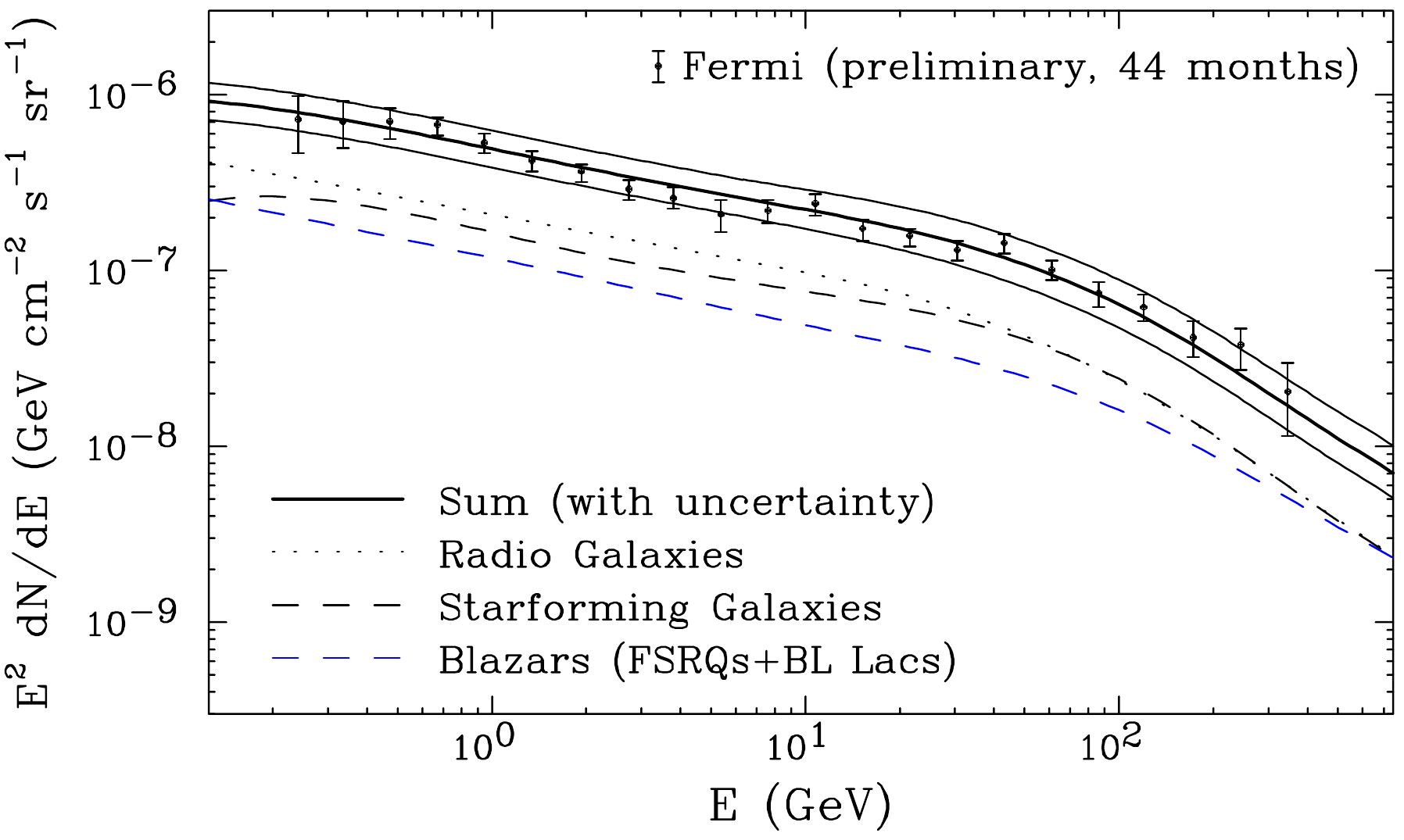}} 
\mbox{\includegraphics[width=0.49\textwidth,clip]{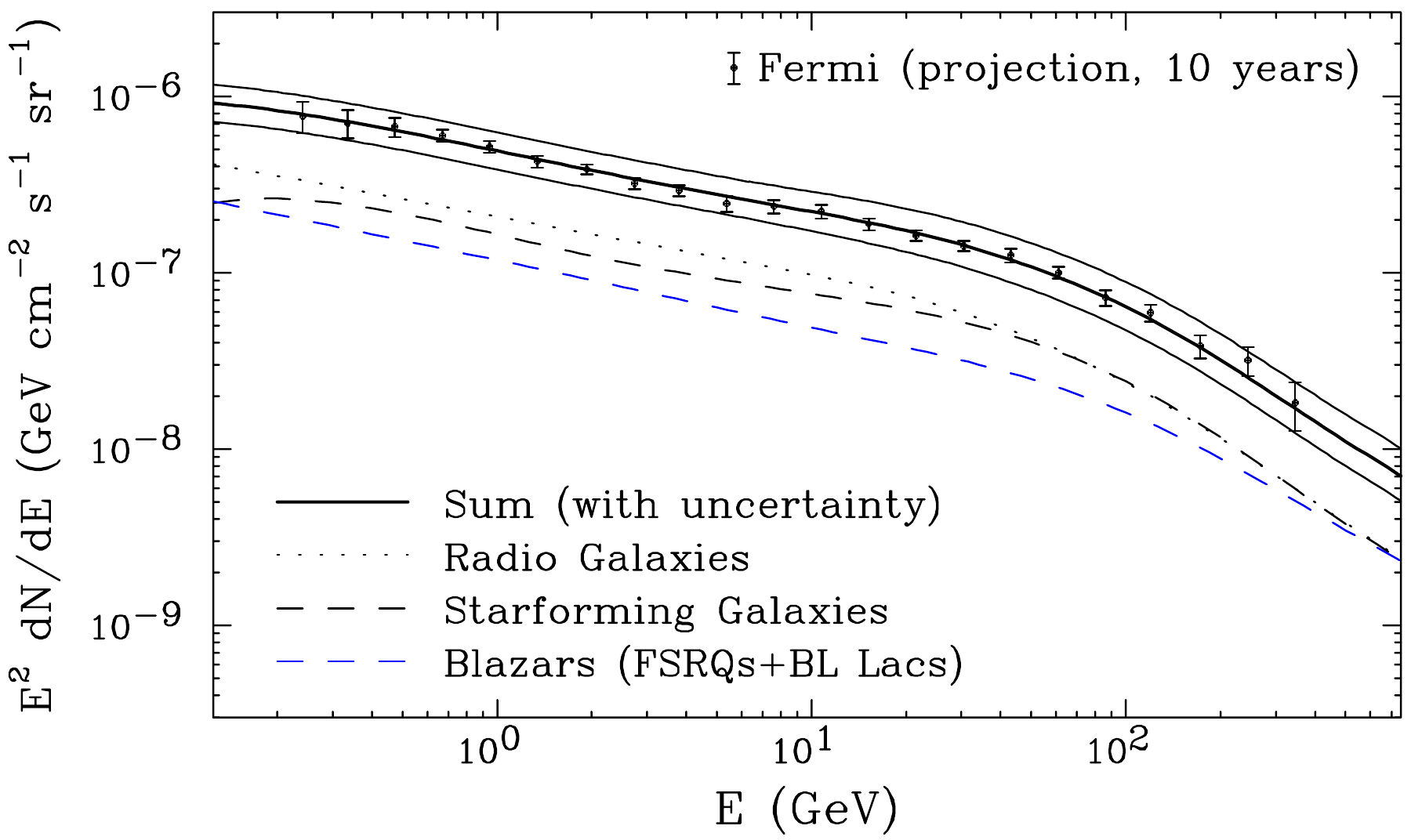}} 
\caption{Projected uncertainties for an astrophysical model of the extragalactic gamma-ray background, after ten years of data from Fermi. In the left frame, we compare this model to the preliminary Fermi measurement~\cite{preliminaryfermi}, whereas in the right frame we compare it to the measurement projected with ten years of data. See text for details.}
\label{projectionsum}
\end{figure*}

To project the error bars on Fermi's future (after 10 total years of operation) measurement of the EGB, we take the preliminary spectrum (which is based on 44 months of data, and is shown in the left frame of Fig.~\ref{projectionsum}~\cite{preliminaryfermi}) and further reduce the size of the error bars by a factor of $\sqrt{120/44}\approx 1.65$. Note that in this projection, we have not removed contributions from to-be-resolved blazars, in order to better facilitate comparisons between projected and current models and measurements.  To project the improvement in the uncertainties of our astrophysical parameters (IR/radio correlation parameters, spectral indices, etc.), we reduce each error bar by the square root of time (relative to the amount of data that was used in the analysis of each source population). We conservatively do not account for any possible improvements in the uncertainties of the radio or IR luminosity functions when making our projections. 
 
\begin{figure}
\mbox{\includegraphics[width=0.49\textwidth,clip]{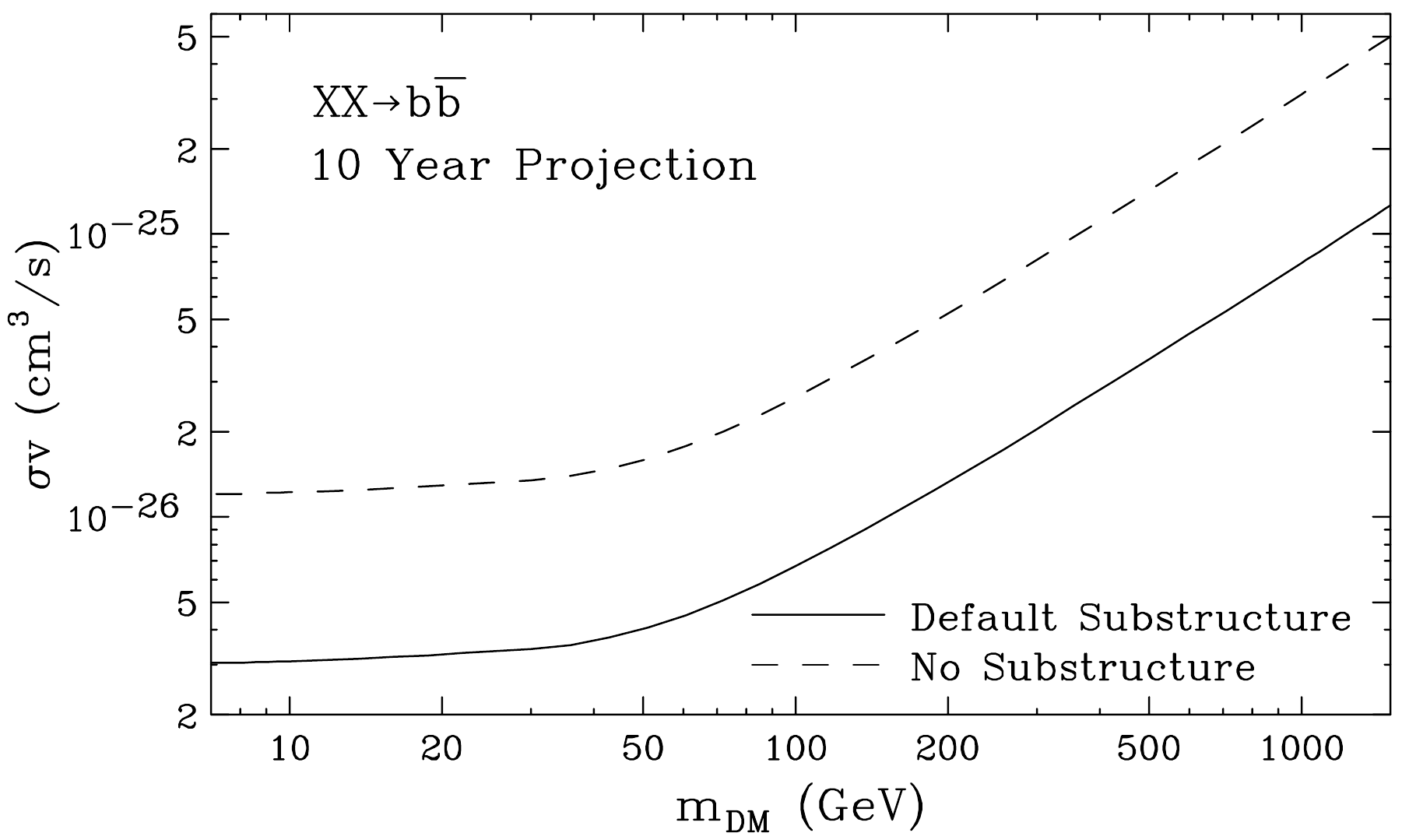}} 
\caption{Our projected sensitivity to dark matter annihilation from Fermi measurements of the EGB after 10 years of operation, using the astrophysical model and projected error bars as shown in the right frame of Fig.~\ref{projectionsum}. See text for details.}
\label{projectedlimits}
\end{figure}

In each frame of Fig.~\ref{projectionsum}, we show the projected uncertainties for an astrophysical model of the EGB after 10 years of Fermi data. In the left frame, we compare this to the preliminary Fermi (44 month) measurement of the EGB~\cite{preliminaryfermi}. In the right frame, we compare this model to our projection for Fermi's measurement of the EGB with 10 years of data. Using this projection for the model parameters and EGB measurements, we repeat the procedure used in Sec.~\ref{constraints} to predict the constraints that Fermi should be able to place on the dark matter annihilation cross section after 10 years of observation. These projected constraints are shown in Fig.~\ref{projectedlimits}.

\section{Summary and Conclusions}
\label{conclusions}

The extragalactic gamma-ray background (EGB) as measured by the Fermi Gamma-Ray Space Telescope contains contributions from a variety of astrophysical sources, including radio galaxies, star-forming galaxies, and blazars.  Fermi observations of individual members of these source classes have been used to construct distribution functions for these populations in both luminosity and redshift. As Fermi collects more data, these distributions will become more tightly constrained, making it possible to determine their contributions to the EGB with increasing precision. 

In this paper, we have constructed a model for the astrophysical contributions to the EGB, and used this model along with Fermi's measurement of the EGB to constrain the contribution from annihilating dark matter. Included in this calculation are contributions from dark matter annihilating in the halos and subhalos distributed throughout the universe, as well as that of the Milky Way's halo and subhalos. The limits on the dark matter's annihilation cross section that we derive in this study are competitive with those based on observations of the Galactic Center and dwarf spheroidal galaxies. In particular, adopting a substructure model based on the extrapolation of numerical simulations (our ``default'' model), the limits presented here are, for all masses, more stringent than those from dwarf galaxies, as recently published by the Fermi collaboration. If we conservatively neglect the contributions from subhalos, our limits become somewhat less stringent (by a factor of $\sim$4-5) but are still competitive with those derived from dwarfs.

As Fermi collects more data, it will not only be capable of measuring the spectrum of the EGB with greater precision, but will also more stringently constrain the characteristics of the various astrophysical source populations that contribute to the EGB.  As a result, we project that Fermi will ultimately be able to achieve a sensitivity to dark matter annihilation products in the EGB that exceeds current constraints by a factor of $\sim$5-10. For our default substructure model, we project that the Fermi measurement of the EGB will ultimately be sensitive to dark matter with the canonical thermal annihilation cross section ($\sigma v = 3\times 10^{-26}$ cm$^3$/s) for masses up to $\sim$400 GeV. 
At the end of Fermi's mission, such limits will likely be the strongest constraints on the dark matter annihilation cross section, although constraints from cosmic-ray observations could in some cases be competitive and complementary~\cite{Cirelli:2013hv, Tavakoli:2013zva,Donato:2008yx, Hailey:2013hxa, Fornengo:2013osa,Bergstrom:2013jra,Evoli:2011id}.

Finally, we stress that dark matter searches making use of the EGB are complementary to those based on observations of the Galactic Center and dwarf galaxies.  The main systematic error in searches involving the region of the Galactic Center arises from uncertainties in the distribution of dark matter in the Milky Way's inner halo.  While the uncertainties faced here regarding dark matter substructure are of a comparable magnitude, they are independent of those issues pertaining to the Inner Galaxy. Furthermore, while the constraints derived from dwarf galaxies are likely more robust to systematic uncertainties than those based on either the Galactic Center or the EGB, they are also somewhat less stringent. As Fermi collects more data, all three of these search techniques will become significantly more powerful, and together will be able to test a wide range of models in which the dark matter consists of thermal relics with masses up to $\sim$400 GeV.

\bigskip
\bigskip

{\it Acknowledgements}: We would like to thank Shin'ichiro Ando and Andrew Hearin for helpful discussion. This work has been supported by the US Department of Energy and by the Kavli Institute for Cosmological Physics. IC and DH would like to thank the Aspen Center for Physics and the NSF Grant 1066293 for hospitality during the earlier stages of this project.

\end{document}